\documentclass[apj]{emulateapj}
\usepackage{graphicx} 
\newcommand{\ha}{H$\alpha$}
\newcommand{\hb}{H$\beta$}
\newcommand{\neiii}{[Ne\thinspace{III}]}
\newcommand{\nii}{[N\thinspace{II}]}
\newcommand{\oii}{[O\thinspace{II}]}
\newcommand{\oiii}{[O\thinspace{III}]}
\newcommand{\sii}{[S\thinspace{II}]}

\begin{document}
\title{Mapping the Youngest Galaxies to Redshift One\altaffilmark{1,2}}
\author{
Yuko~Kakazu,$\!$\altaffilmark{3} \email{kakazu@ifa.hawaii.edu}
Lennox~L.~Cowie,$\!$\altaffilmark{3} \email{cowie@ifa.hawaii.edu}
Esther~M.~Hu,$\!$\altaffilmark{3} \email{hu@ifa.hawaii.edu}
}

\altaffiltext{1}{Based in part on data obtained at the Subaru Telescope,
   which is operated by the National Astronomical Observatory of Japan.}
\altaffiltext{2}{Based in part on data obtained at the W. M. Keck
   Observatory, which is operated as a scientific partnership among the
   the California Institute of Technology, the University of
   California, and NASA and was made possible by the generous financial
   support of the W. M. Keck Foundation.}
\altaffiltext{3}{Institute for Astronomy, University of Hawaii,
   2680 Woodlawn Drive, Honolulu, HI 96822.}

\shorttitle{THE YOUNGEST GALAXIES}
\shortauthors{Kakazu et al.\/}

\slugcomment{Submitted to The Astrophysical Journal}

\begin{abstract}

We describe the results of a narrow band search for ultra-strong
emission line galaxies (USELs) with EW(H$\beta) \ge 30$~\AA. 542
candidate galaxies are found in a half square degree survey
using two $\sim100$\AA\ filters centered at 8150\AA\ and
9140\AA\ with Subaru/SuprimeCam.  Followup spectroscopy has been
obtained for randomly selected objects in the candidate sample
with KeckII/DEIMOS and has shown that they consist of
\oiii$\lambda$5007, \oii$\lambda$3727, and H$\alpha$ selected
strong-emission line galaxies at intermediate redshifts ($z <
1$), and Ly$\alpha$ emitting galaxies at high-redshift ($z >>
5$).  We determine the H$\beta$ luminosity functions and the
star formation density of the USELs, which is 5-10\% of the
value found from ultraviolet continuum objects at $z=0-1$,
suggesting that they correspond to a major epoch in the galaxy
formation process at these redshifts. Many of the USELs show the
temperature-sensitive \oiii$\lambda$4363 auroral lines and about
a dozen have oxygen abundances satisfying the criteria of
eXtremely Metal Poor Galaxies (XMPGs).  These XMPGs are the most
distant known today and our high yield rate of XMPGs suggests
that narrowband method is powerful in finding such populations.
Moreover, the lowest metallicity measured in our sample is
12+log(O/H)=7.06 (6.78$-$7.44), which is close to the minimum
metallicity found in local galaxies, though we need deeper
spectra to minimize the errors. HST/ACS images of several USELs
exhibit widespread morphologies from relatively compact high
surface brightness objects to very diffuse low surface
brightness ones.  The luminosities, metallicities and star
formation rates of USELs are consistent with the strong emitters
being start-up intermediate mass galaxies which will evolve into
more normal galaxies and suggest that galaxies are still forming
in relatively chemically pristine sites at $z<1$.

\end{abstract}


\section{Introduction}
\label{secintro}

The study of low-metallicity galaxies is of considerable
interest for the clues that it can provide about the first
stages of galaxy formation and chemical enrichment.  We would
also like to know if there are any genuinely young galaxies
undergoing their first episodes of star formation at low
redshifts. To date, the most metal-poor systems studied have
been the blue compact emission-line galaxies found in the local
Universe, with systems such as I Zw 18 and SBS 0335-052W
defining the low metallicity boundary with measured 12+log (O/H)
of $\sim7.1-7.2$ (\citealt{sar70,thuan05,izo05}).  
More recently, the Sloan Digital Sky
Survey (SDSS) has yielded additional extremely metal-poor
galaxies (XMPGs) (12+log (O/H) $< 7.65$ or $Z < Z_\sun/12$;
\citealt{knia03, izo06a}).  Despite enormous efforts, only a few
dozen such XMPGs are known, all at redshift $z<0.05$
\citep[e.g.,][]{oey06,izoconf}.

Historically, objective prism surveys have been used to select
emission-line galaxies for low-metallicity studies. (e.g. the
Hamburg QSO Survey \citep{pop96} and its HSS sequel
\citep{ugr99} that discovered HS 2134+0400 \citep{pust06} and
the Kitt Peak International Spectroscopy Survey (KISS;
\citealt{sal00,mel02})).  The advantage of using the objective
prism technique rather than the continuum selection, employed
with the SDSS \citep{knia03} or DEEP2 surveys \citep{hoy05}, is
that they have a higher efficiency  and provide a more uniform
selection.  By comparison, continuum/broad-band surveys have a
very low yield rate (8 new XMPGs and 4 recovered XMPGS from an
analysis of 250,000 spectra over $\sim$3000 deg$^2$ for the SDSS
\citep{knia03}, since low-metallicity populations in
their first outburst have weak continuua and strong emission
lines.

An alternative method of discovering strong emission-line,
low-metallicity galaxies is to use narrowband surveys. Strong
emission-line galaxies have historically been picked up in
high-$z$ Lyman alpha searches
\citep[e.g.,][]{stoc98,smitty,lum5,stern00,tran04,aji06} where
they have been considered contaminants. However, the low
redshift emission line galaxies seen in these surveys are of
great interest in their own right as we shall show in the
present paper. While some spectroscopic studies have been
carried out for low-redshift galaxies selected from narrowband
surveys \citep[e.g.,][]{mai06,ly07}, the small sample sizes have 
precluded any detailed investigation of metallicity and 
identification of a low-metallicity population.

The narrowband method probes to much deeper limits than the
objective prism surveys. This enables probing star-forming
populations out to near redshift $z\sim1$ where the cosmic star
formation rates are considerably higher.  Furthermore the
narrow-band emission-line selection can allow us to assemble
very large samples of strong-emission line objects, with a clean
selection of different line types for the construction of
luminosity functions.

Such a sample allows us to address such questions as whether
there are substantial populations of strong star-forming
galaxies with low metallicities among more massive galaxies.
There has been considerable controversy about the interpretation
of the low metallicity measurements in the blue compact galaxy
samples where the ease with which gas may be ejected in these
dwarf galaxies has complicated the picture
\citep[e.g.,][]{corb06} or, at least, resulted in identifying
low metallicity systems which are not forming their first
generation of stars.  The identification of low metallicity
galaxies -- at the level of the XMPGs -- among massive galaxies
can provides less ambiguous examples of galaxies that are
genuinely `young' and caught in the initial stages of star
formation.  Current efforts to identify low metallicity galaxies
from continuum selected surveys
\citep[e.g.,][]{kob03,lil03,KK04,hoy05} have low-metallicity
thresholds that are higher than this -- about one-third solar
(in O/H).  With a narrow-band selection criterion much larger
emission-line samples including such low metallicity galaxies
can be identified.  With these large samples it is also possible
to determine whether there is an observed lower metallicity
threshold for such galaxies, and to estimate what the
contribution of such strong star-formers might be at these
epochs.

In the present work we use a number of deep, narrow-band images
obtained with the SuprimeCam mosaic CCD camera \citep{supcam} on
the Subaru 8.2-m telescope to find a large sample of extreme
emission-line galaxies.  We first ($\S2$) outline the selection
criteria (magnitude and flux thresholds) for the target fields
resulting in a sample of 542 galaxies, which we call USELs
(Ultra-Strong Emission Liners).  We then describe ($\S3$) the
spectroscopic followups for 161 of these galaxies using
multi-object masks with the DEIMOS spectrograph
{\citep{deimos03}} on the 10-m Keck II telescope.  Sample
spectra for each class of object are shown.  Flux calibration
and equivalent width distributions are presented in $\S4$, and
the resulting measured line ratios are discussed. In $\S5$
luminosity functions are constructed and star formation rates
are estimated for the sample. These galaxies are estimated to
contribute roughly 10\% to the measured star-formation rate
(without extinction corrections) at this epoch.  Analysis of the
metallicities is given in $\S6$.  Their morphologies and
dynamical masses are discussed in $\S7$ and a final summary
discussion is given in $\S8$. We use a standard $H_{0}$ = 70 km
s$^{-1}$ Mpc$^{-1}$, $\Omega_{b}$ = 0.3, $\Omega_{\Lambda}$ = 0.7 
cosmology throughout.

\section{The Narrow Band Selection}
\label{secimaging}

The emission line sample was chosen from a set of narrow band
images obtained with the SuprimeCam camera on the Subaru 8.2-m
telescope.  The data were obtained in a number of runs between
2001 and 2005 under photometric or near photometric conditions.
We used two $\sim$120 \AA\ (FWHM) filters centered at nominal
wavelengths of 8150~\AA\  and 9140~\AA\ in regions of low sky
background between the OH bands.  The nominal specifications for
the Subaru filters may be found at
\url{http://www.naoj.org/Observing/Instruments/
SCam/sensitivity.html} and are also described in \citet{aji03}.
We shall refer to these filters as NB816 and NB912.

  \begin{figure}
  \begin{center}
   \includegraphics[viewport=20 0 446 596,width=2.4in,clip,angle=90,scale=0.95]{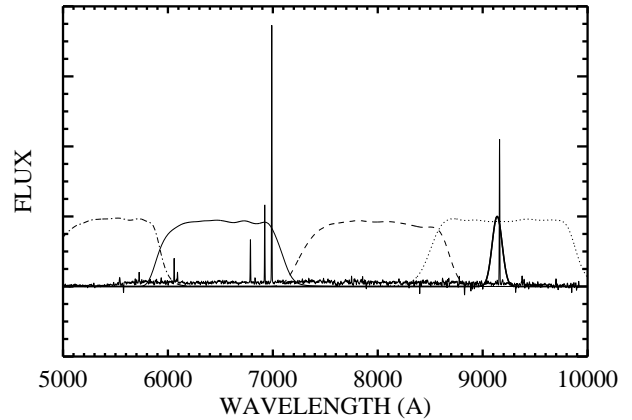}
    \caption{Schematic illustration of the selection process and a 
      typical spectrum of the galaxies we find. The objects are chosen 
      based on their excess light in one of two narrow band filters at 
      8160 \AA\ and 9140\AA.  The present case corresponds to an 
      H$\alpha$ emission line object found in the 9140\AA\ filter 
      (shown with the narrow solid curve). Also illustrated are the 
      broad band $V$ (dash-dot), $R$ (solid), $I$ (dashed), and $z'$ 
      (dotted) filters use to measure the continuum. The spectrum
      shown corresponds to object 205 in Table~\ref{tbl-3} and is an 
      $H\alpha$ emitter at $z=0.3983$. The easily visible lines are the 
      Balmer series and the \protect{\oiii} lines at $\lambda\lambda$5007, 
      4959, and 4363\AA.
  \label{fig1:filt-sel}
   }
  \end{center}
  \end{figure}

 \begin{deluxetable*}{lcccccc}
 \tablecolumns{7}
 \tablecaption{Narrowband Survey Area Coverage}
 \tablewidth{0pt}
 \tablehead{                               
 \colhead{Field}  &  \colhead{RA (J2000)} & \colhead{Dec (J2000)} &
 \colhead{($l^{\rm{II}},b^{\rm{II}}$)} & \colhead{E$_{B-V}$\tablenotemark{a}} & 
 \colhead{NB816}  & \colhead{NB912} \\[0.5ex]
 &     &    &    &   &   \colhead{(arcmin$^{2}$)} & \colhead{(arcmin$^{2}$)}}
 \startdata
 \noalign{\vskip-2pt}
 SSA22 &  22:17:57.00 & +00:14:54.5 &  (\phn{63.1},$-44.1$) &0.07 & 674 & 591\\
 SSA22\_new &  22:18:24.67 & +00:36:53.4 & (\phn{63.6},$-43.9$) & 0.06 & 278 & 278\\
 A370\_new &  02:41:16.27 & $-$01:34:25.1 & (173.4,$-53.3$) & 0.03 & 278 & 278\\
 HDF-N   &  12:36:49.57 & +62:12:54.0 & (125.9,$+54.8$) & 0.01 & 710 & 528 \\
 \noalign{\vskip4pt}
 \cline{6-7}
 \noalign{\vskip6pt}
 Total & & & & & 1940 & 1675\\
 \enddata
 \tablecomments{An adjacent field to A370\_new is a site of a gravitational lensing 
 cluster at $z\sim0.375$, and was omitted from the suvey.}
 \tablenotetext{a\ }{estimated using  \url{http://irsa.ipac.caltech.edu/applications/DUST/} based on \citet{schleg98}}
 \label{tbl-1}
 \end{deluxetable*}

About 5 hour exposures were obtained with NB816 and $\sim$10
hour exposures with NB912 yielding 5 sigma limits fainter than
26 mags in both bands. Deep exposures in $B$, $V$, $R$, $I$ and
$z'$ were also taken for the fields.  The data were taken as a
sequence of dithered background-limited exposures and successive
mosaic sequences were rotated by 90 degrees. Only the central
uniformly covered areas of the images were used. Corresponding
continuum exposures were always obtained in the same observing
run as the narrowband exposures to avoid false identifications
of transients such as high-$z$ supernovae, or Kuiper belt
objects, as emission-line candidates. A detailed description of
the full reduction procedure for images is given in
\citet{capak}.  All magnitudes are given in the AB system
\citep{oke90}. These were measured in $3''$ diameter apertures,
and had average aperture corrections applied to give total
magnitudes.

The primary purpose of the program was to study Ly$\alpha$
emitters at redshifts of $z\sim 5.7$ and $z\sim 6.6$
\citep{lum5,hdf6} but the narrow band imaging is also ideal for
selecting lower redshift emission line galaxies and it is for
this purpose that we use these data in the present paper.  The
fields which we use and the area covered (approximately a half
square degree in each bandpass) are summarized in
Table~\ref{tbl-1}.  These are distributed over the sky to deal with
cosmic variance. We selected galaxies in the narrowband NB816
filter using the Cousins $I$ band filter as a reference
continuum bandpass and including all galaxies with $NB816<25$
and $I-NB816$ greater than 0.8.  We selected galaxies in the
$NB912$ filter with the $z$ filter as the reference continuum
bandpass and included all galaxies with $NB912<25$ and $z-NB912$
greater than 1.  The selection process is illustrated for a
galaxy found in the NB912 filter in Figure~\ref{fig1:filt-sel}.
Both selections correspond roughly to choosing objects with
emission lines with rest frame equivalent widths greater than
100 \AA.  The exact equivalent width limit depends on the
precise position of the emission line in the filter and the
redshift of the galaxy which in turn depends on which emission
line is producing the excess light in the narrow band.

 \begin{figure}
 \begin{centering}
 \includegraphics[viewport=20 0 446 626,width=2.4in,clip,angle=90,scale=0.85]{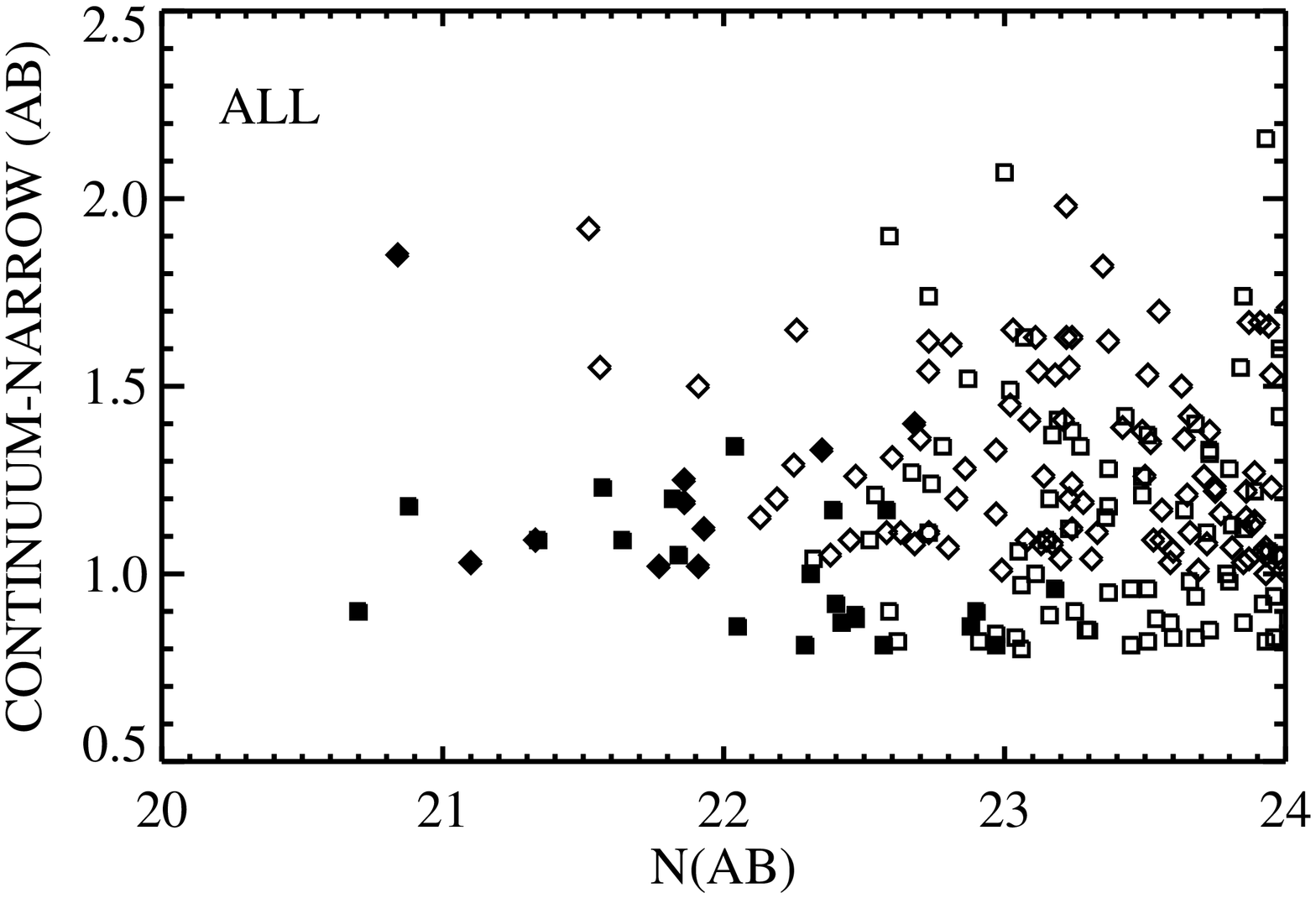}
 \includegraphics[viewport=20 0 446 626,width=2.4in,clip,angle=90,scale=0.85]{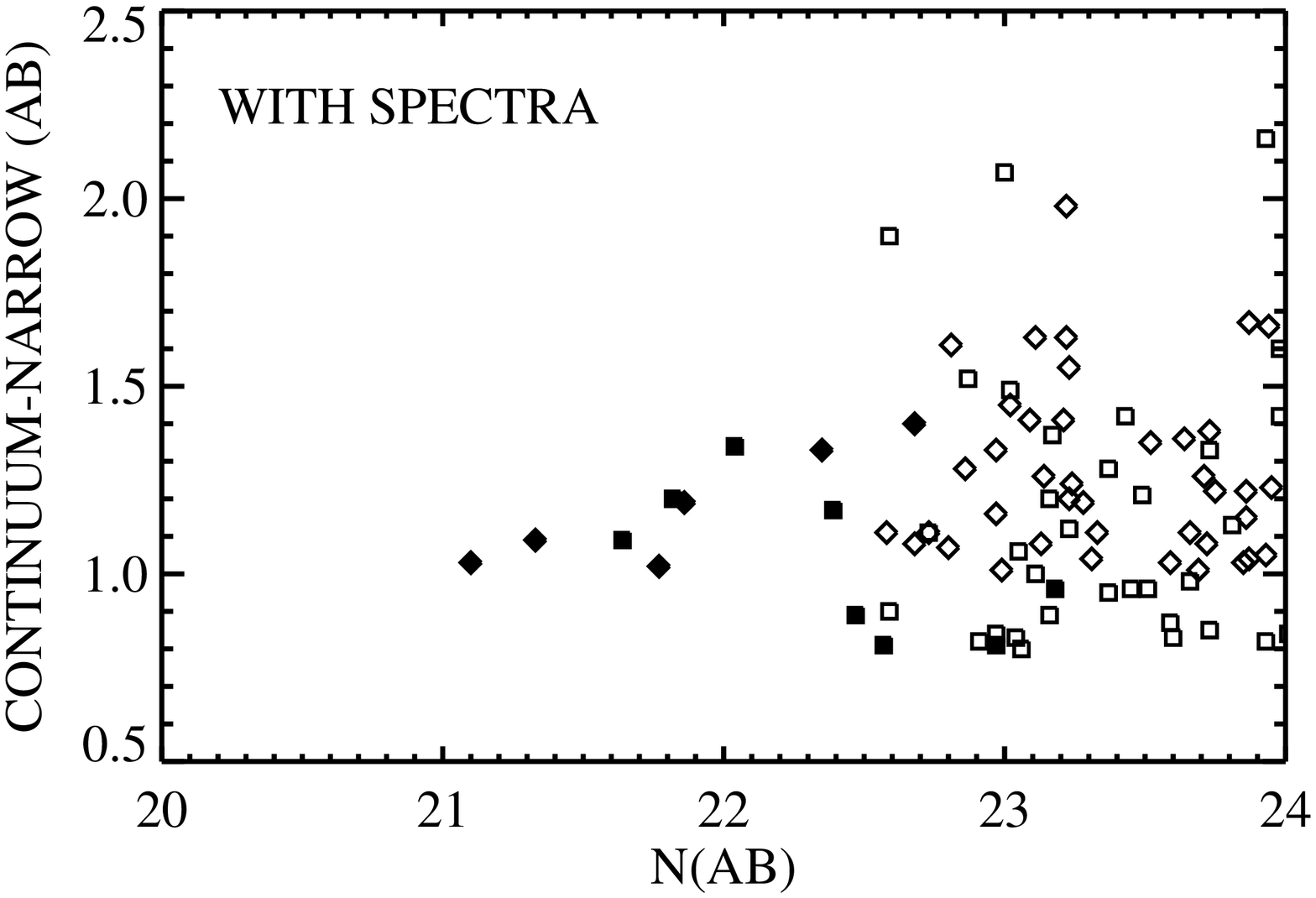}

  \caption{Continuum -- Narrow band magnitude versus narrow
	band magitude for all objects with narrow band magnitude
	brighter than 24. The diamonds show the narrowband excess
        emission magnitude of the NB912 sample
	and the squares the NB816 sample. Galaxies which would
	be included in an $R<24$ continuum selected sample are
	shown with solid symbols. The upper panel shows the
	complete sample while the lower panel shows the subsample
	which has been spectroscopically identified.
  \label{fig0:emit-distrib}
  }
 \end{centering}
 \end{figure}

The final USEL sample consists of 542 galaxies (267 in the NB816
filter and 275 in the NB912 filter).  Tabulated coordinates,
multi-color magnitudes, and redshifts (where measured) for these
objects are summarized in Table~\ref{tbl-2}.  Very few of these
objects would be included in continuum-selected spectroscopic
samples.  Figure~\ref{fig0:emit-distrib} shows the narrow band
excess as a function of narrowband ($N_{AB}$) magnitude for
objects with narrow band magnitudes brighter than 24. The open
symbols show the present sample while the solid symbols show
objects which would be included in an $R<24$ continuum-selected
sample.

\section{Spectra}
\label{secspec}

Spectroscopic observations were obtained for 161 USELs from the sample 
using the Deep Extragalactic Imaging Multi-Object
Spectrograph (DEIMOS; \citealt{deimos03}) on Keck~II in a series of runs
between 2003 and 2006. The emission line objects were included 
in masks designed to observe high-$z$ Ly$\alpha$ candidates
and, as can be seen in the lower panel of Figure~\ref{fig0:emit-distrib},
constitute an essentially random sample of the emission line galaxies.

  \begin{figure*}
  \begin{centering}
  \includegraphics[viewport=20 0 446 616,width=4.8in,clip,angle=90,scale=0.8]{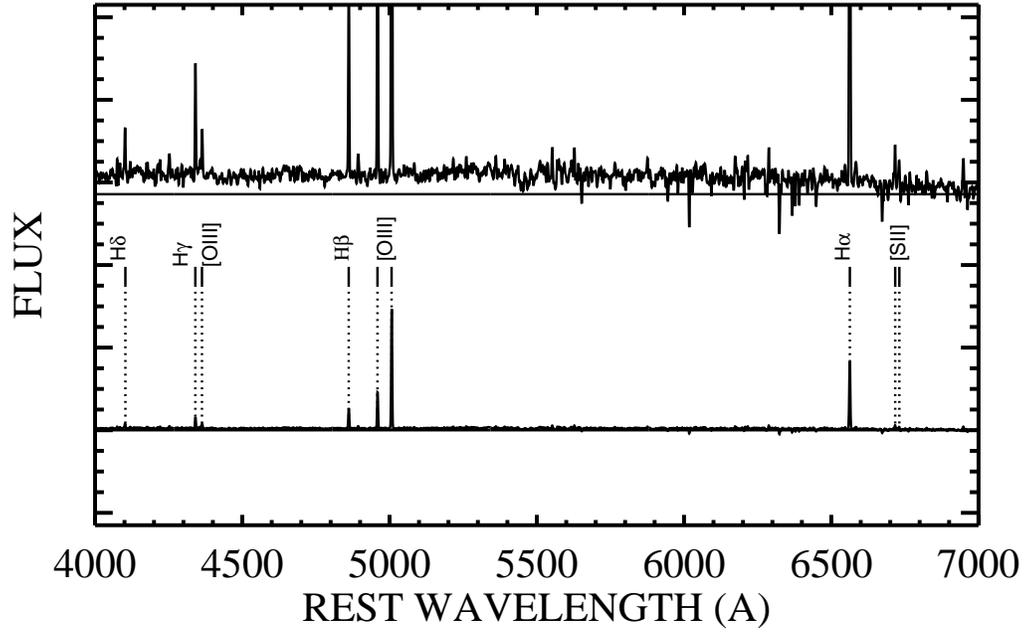}
  \caption{Spectrum of an \protect{\ha} emission galaxy selected 
         in the NB912 filter.  In the upper plot we have decreased the scale
           of the vertical axis by a factor of 10 to show the continuum and the weaker lines. 
The more important emission line features are labelled and marked with the dotted lines.
         \label{fig8:ha-sample}
  }
  \end{centering}
  \end{figure*}

  \begin{figure*}
  \begin{centering}
  \includegraphics[viewport=20 0 446 616,width=4.8in,clip,angle=90,scale=0.8]{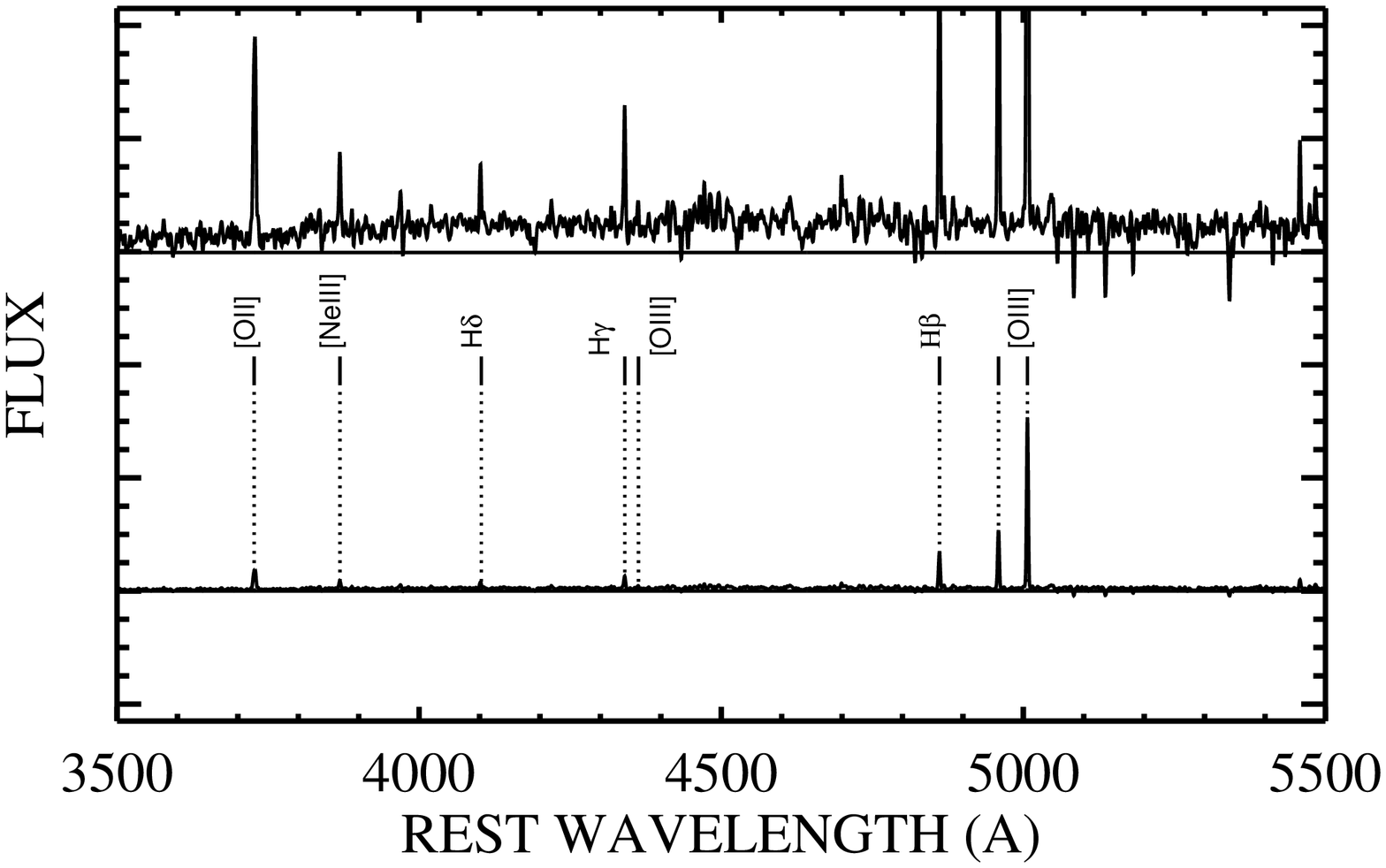}
  \caption{Spectrum of an \protect{\oiii} galaxy in the NB816 selected sample. 
           The lower plot shows the relative strengths of the very 
           strong emission lines in the spectrum. In the upper plot we have decreased
           the scale of the vertical axis by a factor of 10 to show the continuum and the weaker
           lines. The more important emission line features are labelled and marked with the dotted lines.
           \label{fig9:oiii-sample}
  }
  \end{centering}
  \end{figure*}

  \begin{figure*}
  \begin{centering}
  \includegraphics[viewport=20 0 446 616,width=4.8in,clip,angle=90,scale=0.8]{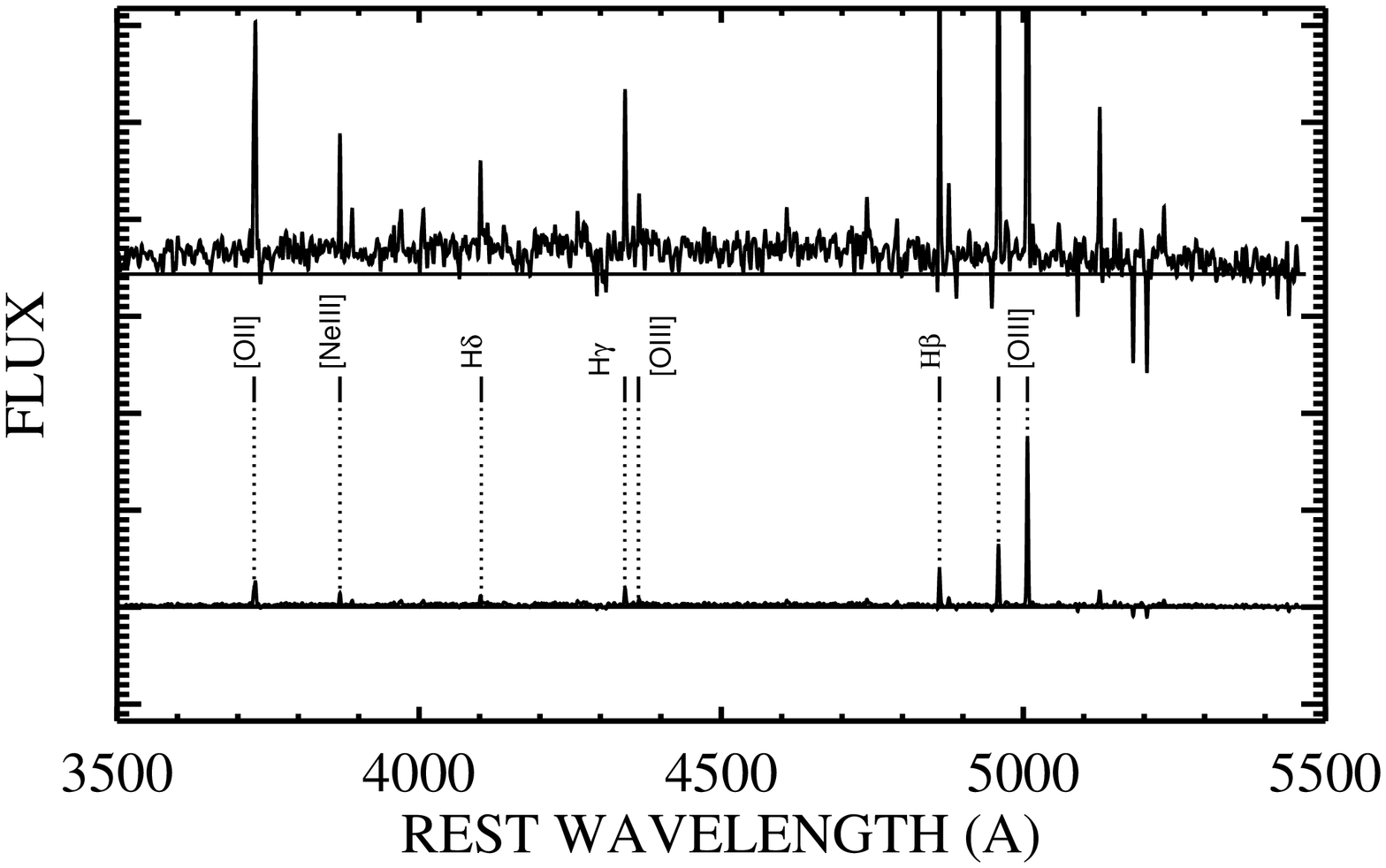}
  \caption{Spectrum of an \protect{\oiii} galaxy selected in the NB912 filter.  
           The lower plot shows the relative strengths of the very 
           strong emission lines in the spectrum. In the upper plot we have decreased
           the scale of the vertical axis by a factor of 10 to show the continuum and the weaker
           lines. The more important emission line features are labelled and marked with the dotted lines.
	   \label{fig10:oiii-sample2}
  }
  \end{centering}
  \end{figure*}

  \begin{figure*}
  \begin{centering}
  \includegraphics[viewport=20 0 446 616,width=4.8in,clip,angle=90,scale=0.8]{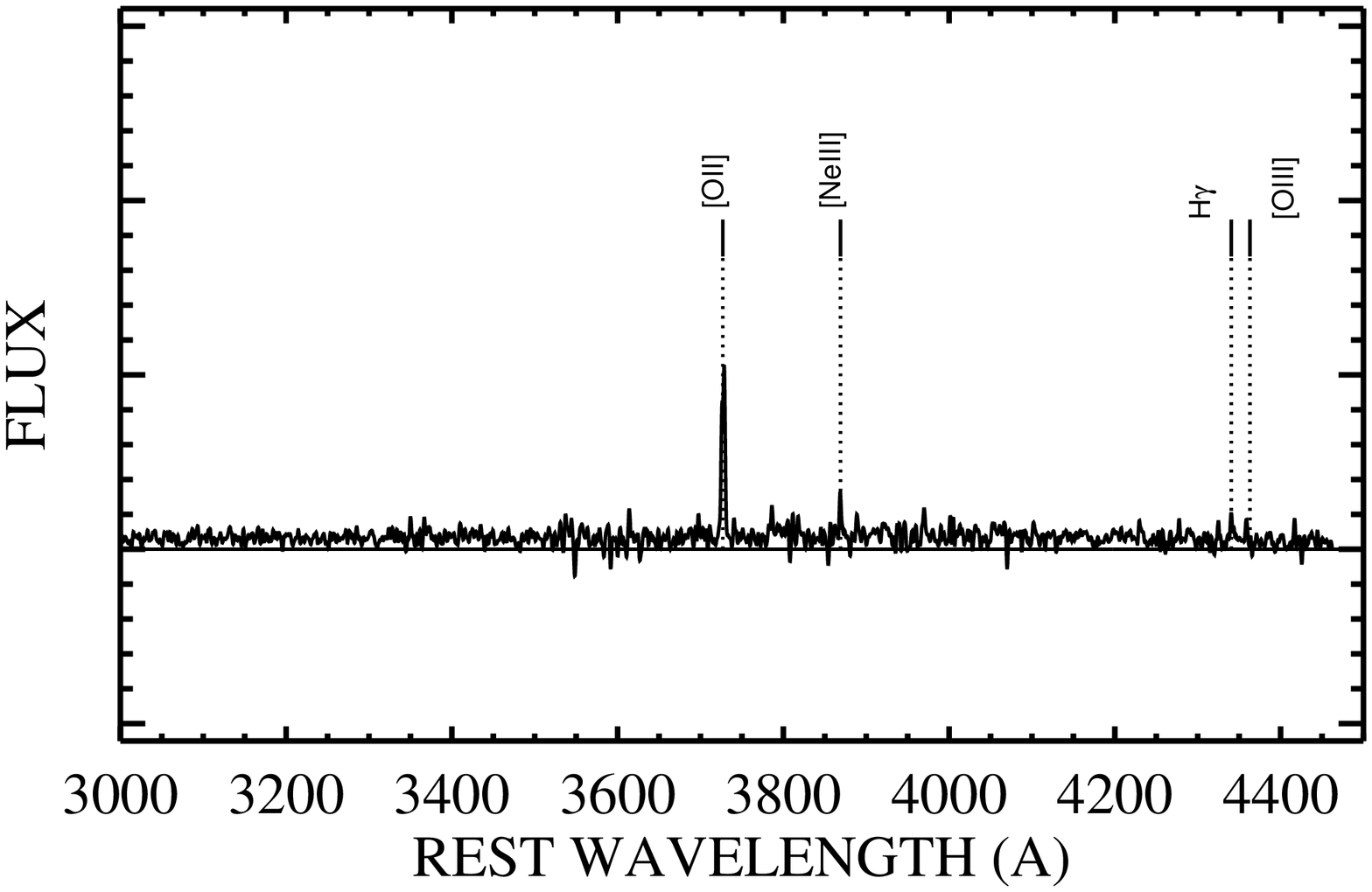}
  \caption{Spectrum of an \protect{\oii} galaxy selected in the NB816 filter.   
           The plot shows the relative strengths of the very 
           strong emission lines in the spectrum. The more important emission line features are labelled and marked with the dotted lines.
	   \label{fig11:oii-sample} 
  }
  \end{centering}
  \end{figure*}

The observations were primarily made with the G830 $\ell$/mm
grating blazed at 8640 \AA\ and used $1''$ wide slitlets.  In
this configuration, the resolution is 3.3 \AA, which is
sufficient to distinguish the \oii $\lambda$3727 doublet
structure. This allows us to easily identify \oii $\lambda$3727
emitters where often the \oii $\lambda$3727 doublet is the only
emission feature.  The spectra cover a wavelength range of
approximately 4000~\AA\ and were centered at an average
wavelength of $7800$~\AA, though the exact wavelength range for
each spectrum depends on the slit position with respect to the
center of the mask along the dispersion direction.  The G830
grating used with the OG550 blocker gives a throughput greater
than 20\% for most of this range, and $\sim28\%$ at 8150 \AA.
The observations were not generally taken at the parallactic
angle, since this was determined by the mask orientation, so
considerable care must be taken in measuring line fluxes as we
discuss below.  Each $\sim 1$~hr exposure was broken into three
subsets, with the objects stepped along the slit by $1.5''$ in
each direction. Some USELs were observed multiple times,
resulting in total exposure times for these galaxies of $2 - 3$
hours.  The two-dimensional spectra were reduced following the
procedure described in \citet{cowie96} and the final
one-dimensional spectra were extracted using a profile weighting
based on the strongest emission line in the spectrum. A small
number of the spectra were obtained with the ZD600 $\ell$/mm
grating giving a correspondingly lower resolution but a wider
wavelength coverage.  These observations were centered at
$7200$~\AA. 

  \begin{figure}
  \begin{centering}
  \includegraphics[viewport=20 0 456 650,width=2.8in,clip,angle=90,scale=0.85]{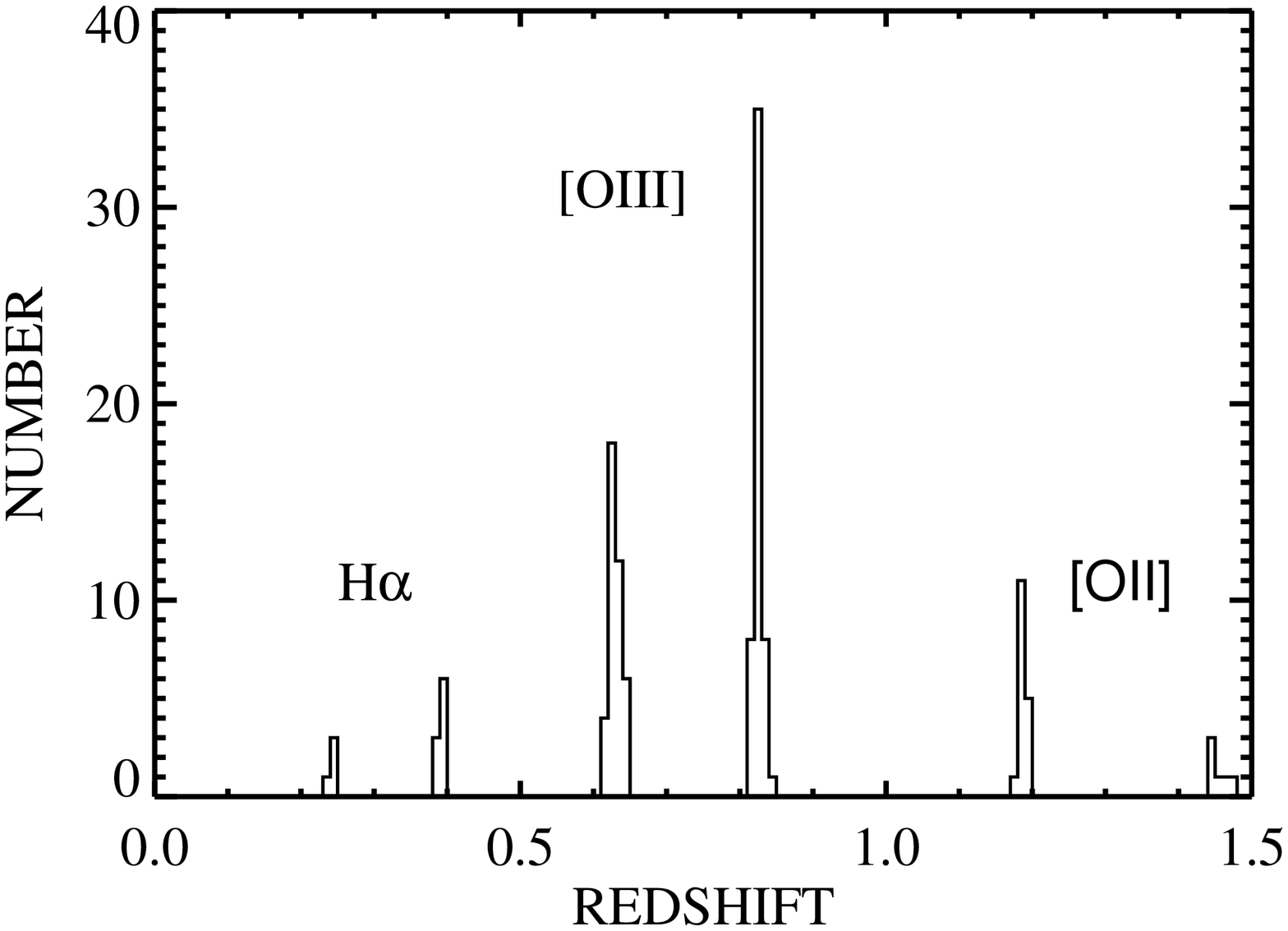}
  \includegraphics[viewport=20 0 456 650,width=2.8in,clip,angle=90,scale=0.85]{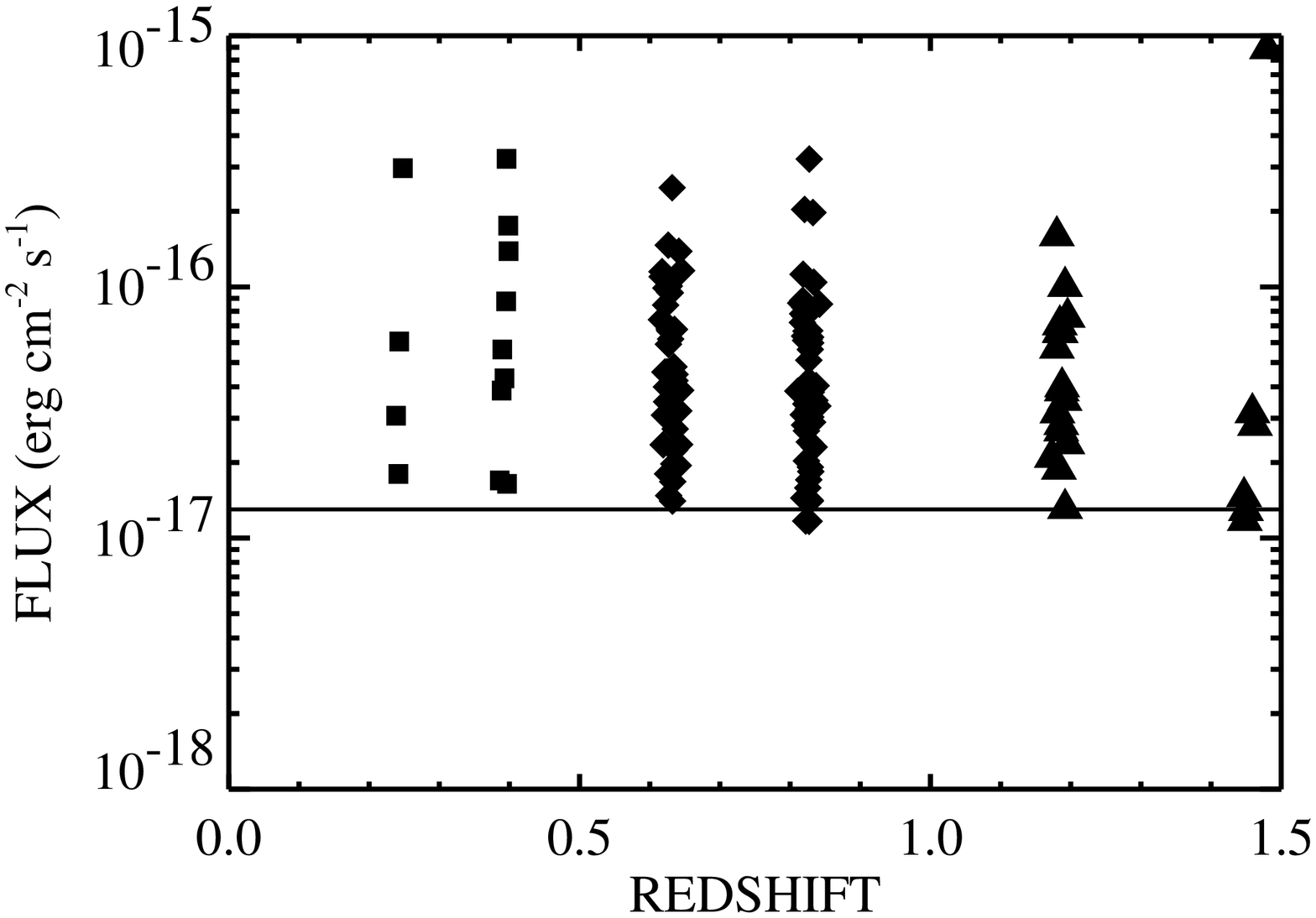}
  \caption{(a) Distribution of redshifts for the spectroscopically identified 
               sources. \protect{\oiii} $\lambda$5007 emitters are the most common.
               Since the focus of this paper is on intermediate-redshift ($z \lesssim 1$)
               strong emission line galaxies, we did not plot high redshift Ly$\alpha$ galaxies ($z >> 5$) 
               in our NEO sample. High-$z$ Ly$\alpha$ emitters are discussed in \citet{lum5,hdf6}.
           (b) Flux versus redhift for the spectroscopically identified sample. 
               Squares are \protect{\ha}, diamonds are  \protect{\oiii} $\lambda$5007,
               and triangles are  \protect{\oii} $\lambda$3727. The solid line shows 
               the flux limit corresponding to the narrow band magnitude limit
               of N(AB)=25 for an emitter with very large equivalent width. Some 
               objects with lower equivalent widths fall below this limit.
               \label{fig7:emit-zdistrib}
    }
   \end{centering}
   \end{figure}

Essentially all of the emission line candidates which were
observed were identified, though two of the objects in the NB816
sample are stars where the absorption line structure mimics
emission in the band. Sample spectra are shown in 
Figures~\ref{fig8:ha-sample}, \ref{fig9:oiii-sample},
\ref{fig10:oiii-sample2}, and \ref{fig11:oii-sample}. The
measured redshifts are given in Tables~\ref{tbl-2} and
\ref{tbl-3}.  The narrow band emission line selection produces a
mixture of objects corresponding to H$\alpha$,
\oiii$\lambda$5007, and \oii$\lambda$3727 and, at the faintest
magnitudes ($>24$), high redshift Ly$\alpha$ emitters.  The
number of objects seen in each line and the redshifts where they
are found are shown in Figure~\ref{fig7:emit-zdistrib}.  The
spectroscopically identified sample from both bands contains 13
H$\alpha$, 92 \oiii$\lambda$5007, and 23 \oii$\lambda$3727
emitters. In the remainder of the paper we shall focus primarily
on the H$\alpha$ and  \oiii$\lambda$5007 selected galaxies which
lie between redshifts zero and one.

Since only 30\% of the USELs are spectroscopically identified we
must apply a substantial incompleteness correction in computing
the line luminosity function and the universal star formation
histories. Because the type mix may vary as a function of
magnitude we have adopted a magnitude dependent weighting for
each galaxy equal to the total number of galaxies at this
magnitude divided by the number of spectroscopically identified
galaxies. However, the analysis is not particularly sensitive to
the adopted scheme since the fraction of identified galaxies is
relatively constant with magnitude.

\section{Flux Calibrations}
\label{secflux}

Generally our spectra were not obtained at the parallactic angle
since this is determined by the DEIMOS mask orientation.
Therefore flux calibration using standard stars is problematic
due to atmospheric refraction effects, and special care must be
taken for the flux calibration.  We thus employed three
independent methods for the flux calibration.  In \S4.1 we
introduce ``primary fluxes'' which are absolute fluxes of the
emission lines used to select the galaxies. Primary fluxes are
computed from the SuprimeCam broadband and narrowband
magnitudes. We use these fluxes to derive luminosity functions
of H$\alpha$ and \oiii$\lambda$5007 emitters at $z < 1$
(\S5.1).

In \S4.2 we measure line fluxes from the spectra.  Relative line
fluxes can be measured from the spectra without flux calibration
as long as we restrict the line measurments to over short
wavelength range where the DEIMOS response is essentially
constant. For example, one can assume the response of
neighboring lines (e.g. \oiii$\lambda$4949 and
\oiii$\lambda$5007) are the same and therefore one can measure
the flux ratio without calibration.  For bright galaxies, we can
absolutely calibrate the fluxes by integrating spectra and
equating them to Subaru broadband fluxes.  These line fluxes
derived from the spectra are used as a check of the primary
fluxes.  We show that the ratio of
\oiii$\lambda$5007/\oiii$\lambda$4959 is indeed close to 1/3,
and that the fluxes computed from the spectra are highly
consistent with the primary fluxes measured in \S4.1. In \S4.3,
we show Balmer flux ratios f(H$\beta$)/f(H$\alpha$) of bright
H$\alpha$ emitters are close to the Case B conditions,
suggesting very little reddening.

Metallicity measurements by the direct method require four
emission lines that are widely displaced over the spectral
wavelength range (\oiii$\lambda$$\lambda$4959, 5007,
\oiii$\lambda$4363, and \oii$\lambda$3727).  To calibrate these
lines, we used neighboring Balmer lines with the assumption of
Case B conditions, and this is described in \S4.4.

\subsection{Narrow Band Fluxes $-$ {\it Primary Fluxes} }
\label{secnarrow}

For the emission lines used to select each galaxy we may compute
the equivalent widths and absolute fluxes directly from the
narrow band magnitudes (N) and the corresponding continuum
magnitudes (C) from our SuprimeCam imaging data.  For example,
in the case of the NB816 selected emission-line galaxies, N
corresponds to the NB816 magnitude and C is the $I$ band
magnitude.  We shall refer to the values calculated in this way
as the {\it primary fluxes} and use this quantity to compute the
luminosity functions for each emitter in \S5.1.

Defining the quantity

\begin{displaymath}
        R=10^{-0.4*(N-C)}
        \label{eq:1}
\end{displaymath}
the observed frame equivalent width becomes
\begin{displaymath}
        EW = \left [ {R - 1} \over {\displaystyle {\phi - 
             {R\over{\Delta\lambda}}}} \right ]
        \label{eq:2}
\end{displaymath}
where $\phi$ is the narrow band filter response normalized such
that the integral over wavelength is unity and $\Delta\lambda$
is the effective width of the continuum filter.  The narrow band
filter is often assumed to be rectangular in which case $\phi$
becomes $1/\delta\lambda$ where $\delta\lambda$ is the width of
the narrow band but as can be seen from Figure~\ref{fig1:filt-sel} 
this is not a very good approximation in the present case.  For
very high equivalent width objects the denominator in this
equation becomes uncertain unless the broad band data are very
deep, and this can result in a large scatter in the very highest
equivalent widths where the continuum is poorly determined.

In the case of the \oiii$\lambda$5007 line we must include the
second member of the doublet which also lies within the narrow
band filter. We have computed these cases assuming the flux of
the \oiii$\lambda$4959 line is 0.34 times that of the
\oiii$\lambda$5007 line. Then $\phi=\phi_1 + 0.34 \times \phi_2$
where $\phi_1$ is the filter response at the redshifted
5007~\AA\ wavelength and $\phi_2$ is the filter response at
redshifted 4959~\AA.

  \begin{figure}
  \begin{centering}
  \includegraphics[viewport=20 0 456 660,width=2.2in,clip,angle=90,scale=0.8]{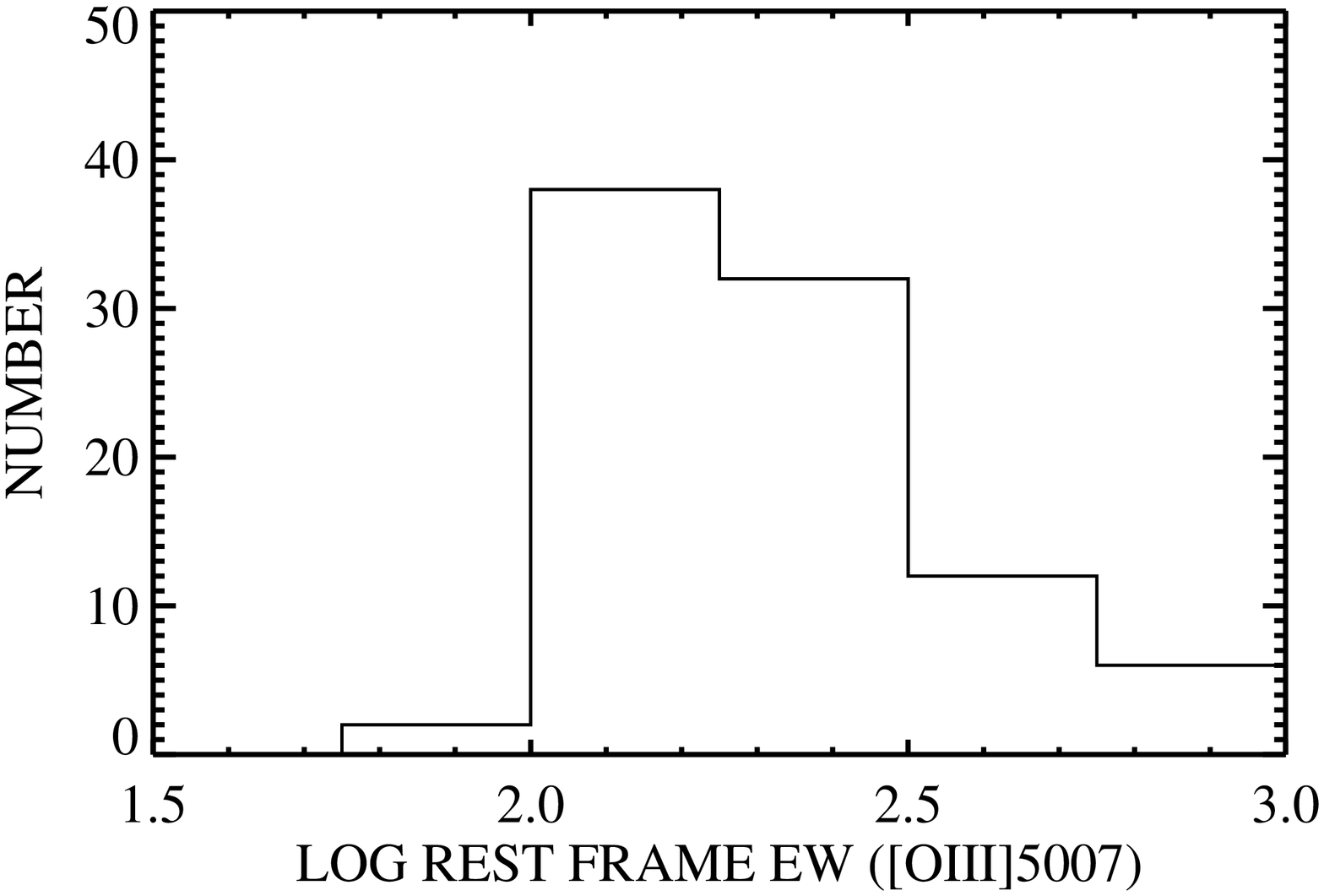}
  \includegraphics[viewport=20 0 456 660,width=2.2in,clip,angle=90,scale=0.8]{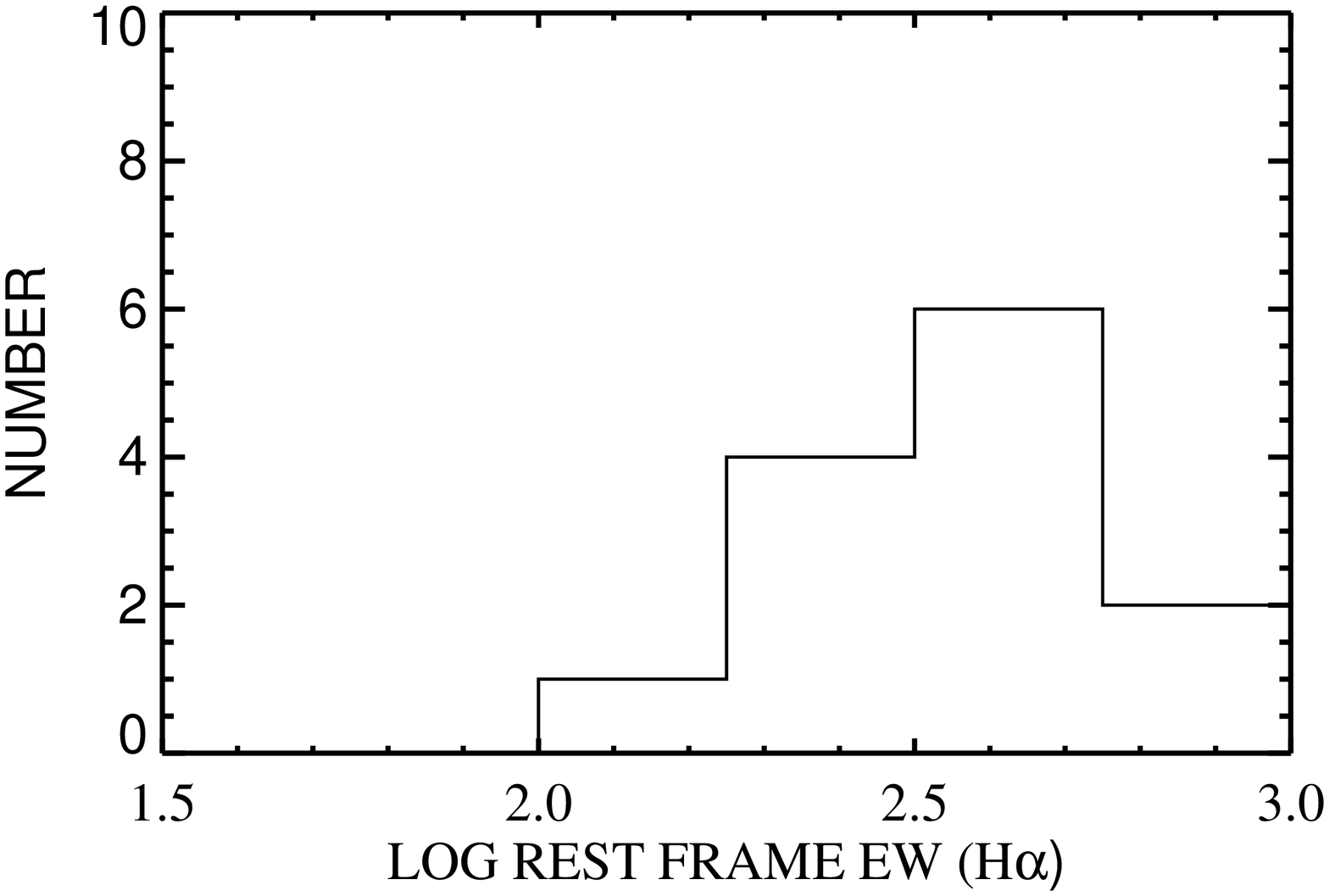}
  \caption{(a) Distribution of the rest frame equivalent widths determined
       from the narrow band magnitudes for the spectroscopically
       identified \protect{\oiii} $\lambda$5007 sources. 
       (b) Distribution of the rest frame equivalent widths for the H$\alpha$
       selected sample. 
  \label{figx:ew-distrib}
  }
  \end{centering}
  \end{figure}

The distribution of the rest frame equivalent widths 
for the H$\alpha$ and \oiii$\lambda$5007 samples is shown in
Figure~\ref{figx:ew-distrib}. The \oiii$\lambda$5007 sample
selects objects with rest frame equivalent widths
above about 100\AA\ while the lower redshift H$\alpha$
sample selects objects with rest frame equivalent widths
above about 150\AA. Since the \oiii$\lambda$5007 lines are
also generally stronger than the H$\alpha$ lines the
\oiii\ selection chooses less extreme objects than the
H$\alpha$ selection and will include a larger fraction
of galaxies at the given redshift.

The high observed frame equivalent widths make the line
fluxes insensitive to the continuum determination and these may
simply be found from

\begin{displaymath}
        f = A \left
[ {10^{-0.4N}} - {10^{-0.4C}} \over {\displaystyle {\phi - 
             {1\over{\Delta\lambda}}}} \right ]
        \label{eq:3}
\end{displaymath}
where A is the AB zeropoint at the narrow band wavelength in
units of erg cm$^{-2}$ s$^{-1}$ \AA$^{-1}$.  The flux depends on
the filter response at the emission line wavelength and
correspondingly is most uncertain at the edges of the filters
where this quantity changes rapidly.  Primary fluxes defined
here are measured by using narrowband (N) and broadband (C)
magnitudes from Subaru imaging data with the object redshift
information from Keck spectra for the filter response at the
exact location of emission line wavelength ($\phi$).  We use
these primary fluxes to construct the luminosity functions of
H$\alpha$ and \oiii$\lambda$5007 selected emitters as we discuss
in \S5.1.

\subsection{Line Fluxes from the Spectra}
\label{secline}

For the short wavelength range where DEIMOS response is
essentially constant, we may also compute the relative line
fluxes from the spectra without calibration.  For each spectrum
we fitted a standard set of lines. For the stronger lines we
used a full Gaussian fit together with a linear fit to the
continuum baseline.  For weaker lines we held the full width
constant using the value measured in the stronger lines and set
the central wavelength to the nominal redshifted value.  We also
measured the noise as a function of wavelength by fitting to
random positions in the spectrum and computing the dispersion in
the results.

  \begin{figure}
  \begin{centering}
  \includegraphics[viewport=20 0 446 660,width=2.8in,clip,angle=90,scale=0.8]{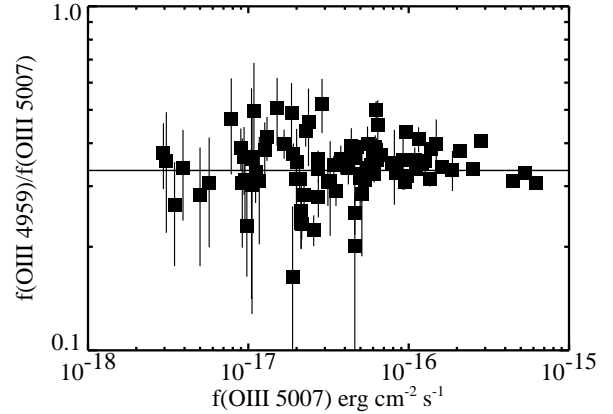}
  \caption{The ratio of the \protect{\oiii} $\lambda$4959 line to 
	  \protect{\oiii} $\lambda$5007. The errors are plus and minus
	  1 sigma. The median ratio is 0.338 and the scatter
	  around this value is consistent with that expected
	  from the statistical errors.
	  \label{fig4:line-ratios}
  }
  \end{centering}
  \end{figure}

These fits should provide accurate relative fluxes over short
wavelength intervals where the DEIMOS response is similar, but
may be expected to be poorer over longer wavelength intervals
where the true calibration can vary from the adopted value.  We
tested the relative flux calibration for neighboring lines and
the noise measurement by measuring the ratio of the \oiii
$\lambda$4959/ \oiii$\lambda$5007 lines. This is expected to
have a value of approximately 0.34.  The ratio is shown as a
function of the \oiii $\lambda$5007 flux in
Figure~\ref{fig4:line-ratios}.  The measured values scatter
around the expected value and the dipsersion is consistent with
the noise determination.  This result supports our assumption of
\oiii$\lambda$4959/\oiii$\lambda$5007 = 0.34 in the primary
fluxes measurements described in \S4.1.

  \begin{figure}
  \begin{centering}
  \includegraphics[viewport=20 0 446 616,width=2.8in,clip,angle=90,scale=0.8]{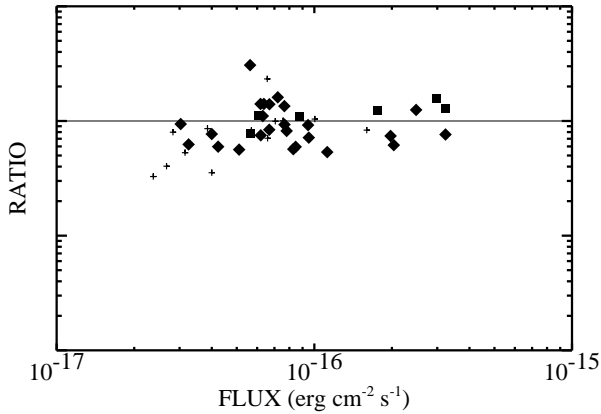}
  \caption{
         Ratio of fluxes computed from the spectra and the
         broad band magnitudes versus those from the narrow
         band magnitudes. \protect{\ha} lines are shown as solid
         boxes, \protect{\oiii} $\lambda$5007 lines as diamonds and 
	 \protect{\oii} $\lambda$3727 lines as crosses.
	 \label{fig3:flux-ratios}
  }
  \end{centering}
  \end{figure}

The brighter objects may be absolutely calibrated 
using the continuum magnitudes obtained from our Subaru data.
We integrated the spectrum convolved with each SuprimeCam filter
response and set this equal to the broad band flux to normalize
the spectrum in each of the filters. We then used the Gaussian
fits to obtain the spectral line fluxes for lines lying within
that broad band. This procedure only works for sources with well
determined continuum magnitudes ($C<24.5$) where the sky
subtraction can be well determined in the spectra. For these
objects the spectrally determined fluxes are shown versus the
primary fluxes in Figure~\ref{fig3:flux-ratios} where we plot
the ratio of the spectral to the primary flux versus the primary
line flux.  The values agree extremely well though the measured
spectral line fluxes are on average about $80-90$\% of the
primary flux values. This may reflect a selection bias in the
choice of the objects or the narrow band filters could be
slightly narrower than the nominal profiles.

\subsection{Balmer Ratios}
\label{secbr}

  \begin{figure}
  \begin{centering}
  \includegraphics[viewport=20 0 446 660,width=2.8in,clip,angle=90,scale=0.8]{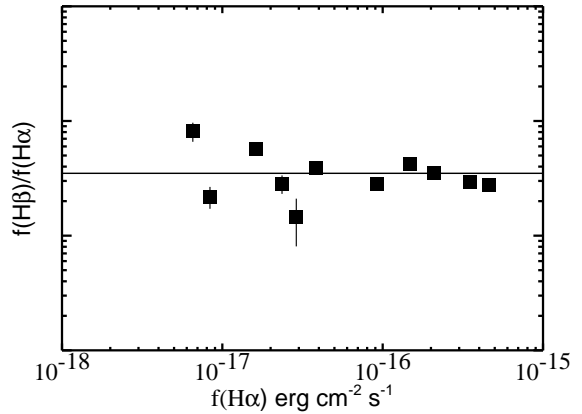}
  \caption{The ratio of the \protect{\hb/\ha} 
	  fluxes versus H$\alpha$ flux. The values average to the unreddened Balmer
	  decrement shown by the solid line but at the lower
	  fluxes the scatter is larger than expected from the
	  statistical errors reflecting the calibration uncertainties
	  for the fainter sources. The figure shows the ten objects detected
          in the H$\alpha$ line with continuum magnitudes above 24.5 in
          the bandpasses corresponding to the lines.
	  \label{fign:line-ratios1}
  }
  \end{centering}
  \end{figure}

We now measured the Balmer ratios for the sample of objects
selected in H$\alpha$ and where the continuum magnitudes were
bright enough to absolutely flux calibrate the spectra.  The
ratio of f(H$\beta$)/f(H$\alpha$) is shown in
Figure~\ref{fign:line-ratios1}. The values average closely to
the Case B ratio which is shown as the solid line and at
brighter fluxes the individual values also closely match to this
value suggesting that the galaxies have very little reddening.
However, at fainter fluxes the scatter about the average value
is considerably higher than the statistical errors. At the
faintest fluxes it appears that the systematic uncertainty can
be as high as a multiplicative factor of two.

\subsection{Final Flux Calibration for Metallicity Analysis}
\label{secff}

For the metallicity analysis we adopted the procedure of
normalizing the longer wavelength lines
(\oiii$\lambda\lambda$4959, 5007, \oiii$\lambda$4363) to their
nearest Balmer line to determine the unreddened fluxes.  For
example, in the case of the H$\alpha$ emission selected
galaxies, we can measure H$\alpha$ absolute fluxes by the
primary fluxes method described in \S4.1. We can then derive
H$\beta$ and H$\gamma$ fluxes from H$\alpha$ fluxes by assuming
Case B recombination [e.g., f(H$\alpha$) = 2.85 $\times$
f(H$\beta$), f(H$\gamma$) = 0.469 $\times$ f(H$\beta$) at T =
$10^4$ K and $N_e \sim 10^2 - 10^4 {\rm cm^{-3}}$;
\citealt{oster}].  As H$\beta$ and \oiii$\lambda\lambda$4959,
5007 have very similar DEIMOS response, the relative fluxes
should remain the same with or without the flux calibration and
this can be expressed by a simple equation:

\begin{displaymath}
   {f_{0}({\rm H}\beta) \over {f_{0}({\rm \oiii}\lambda 4959, \lambda5007)}} =
   {f({\rm H}\beta) \over {f({\rm \oiii}\lambda 4959, \lambda5007)}}
   \label{eq:4}
\end{displaymath}
where $f_{0}$(H$\beta$) and $f_{0}$(\oiii$\lambda$4959,
$\lambda$5007) are the flux counts in the un-calibrated,
reddened DEIMOS spectra, while f(H$\beta$) and
f(\oiii$\lambda$4959, $\lambda$5007) are absolute, unreddened
fluxes.  Since we know f(H$\beta$) from f(H$\alpha$) with the
Case B assumption and
$f_{0}$(H$\beta$)/$f_{0}$(\oiii$\lambda$4959, $\lambda$5007)
from the DEIMOS spectra, we can derive f(\oiii$\lambda$ 4959,
$\lambda$5007) using this simple formula.  Similary, we can
absolutely calibrate \oiii$\lambda$4363 lines by using its
neighboring Balmer line, H$\gamma$:

\begin{displaymath}
   {f_{0}({\rm H}\gamma) \over {f_{0}({\rm \oiii}\lambda 4363)}} =
   {f({\rm H}\gamma) \over {f({\rm \oiii}\lambda 4363)}}
   \label{eq:5}
\end{displaymath}
where $f_{0}$(H$\gamma$) and $f_{0}$(\oiii $\lambda$4363) are
again the counts in flux uncalibrated, reddened DEIMOS spectra,
and $f$(H$\gamma$) and $f$(\oiii $\lambda$4363) are absolute
fluxes.

In the case of the \oiii\ selected emitters, we first derive
\oiii $\lambda\lambda$4959, 5007 absolute fluxes using the
primary fluxes method (\S4.1), and then use the above formula to
get absolute fluxes of H$\beta$, then H$\gamma$ (by the Case B
ratio), and finally \oiii$\lambda$4363.

This flux calibration technique using neighboring Balmer line
should work well for the \oiii$\lambda\lambda$4959, 5007,
$\lambda$4363 lines and the \nii\ lines which all lie close to
Balmer lines but may be slightly more approximate for the
\sii\ lines. The higher order Balmer lines are too uncertain to
apply this procedure due to inadequate S/N of the lines, and we
have used the continuum flux calibrated values with no reddening
for the \oii$\lambda$3727 and \neiii\ lines. These values will
have correspondingly higher flux uncertainties. Fortunately the
\oii$\lambda$3727 line is very weak in most of the objects and
the uncertainty has little effect on the metallicity
determinations. However, ionization analyses based on the
\neiii\ line should be undertaken with caution.

\section{Star Formation History}
\label{secsfr}

\subsection{H$\alpha$ and \oiii$\lambda$5007 Luminosity Functions}
\label{seclf}

  \begin{figure}
\includegraphics[viewport=20 0 460 646,width=2.8in,clip,angle=90,scale=0.85]{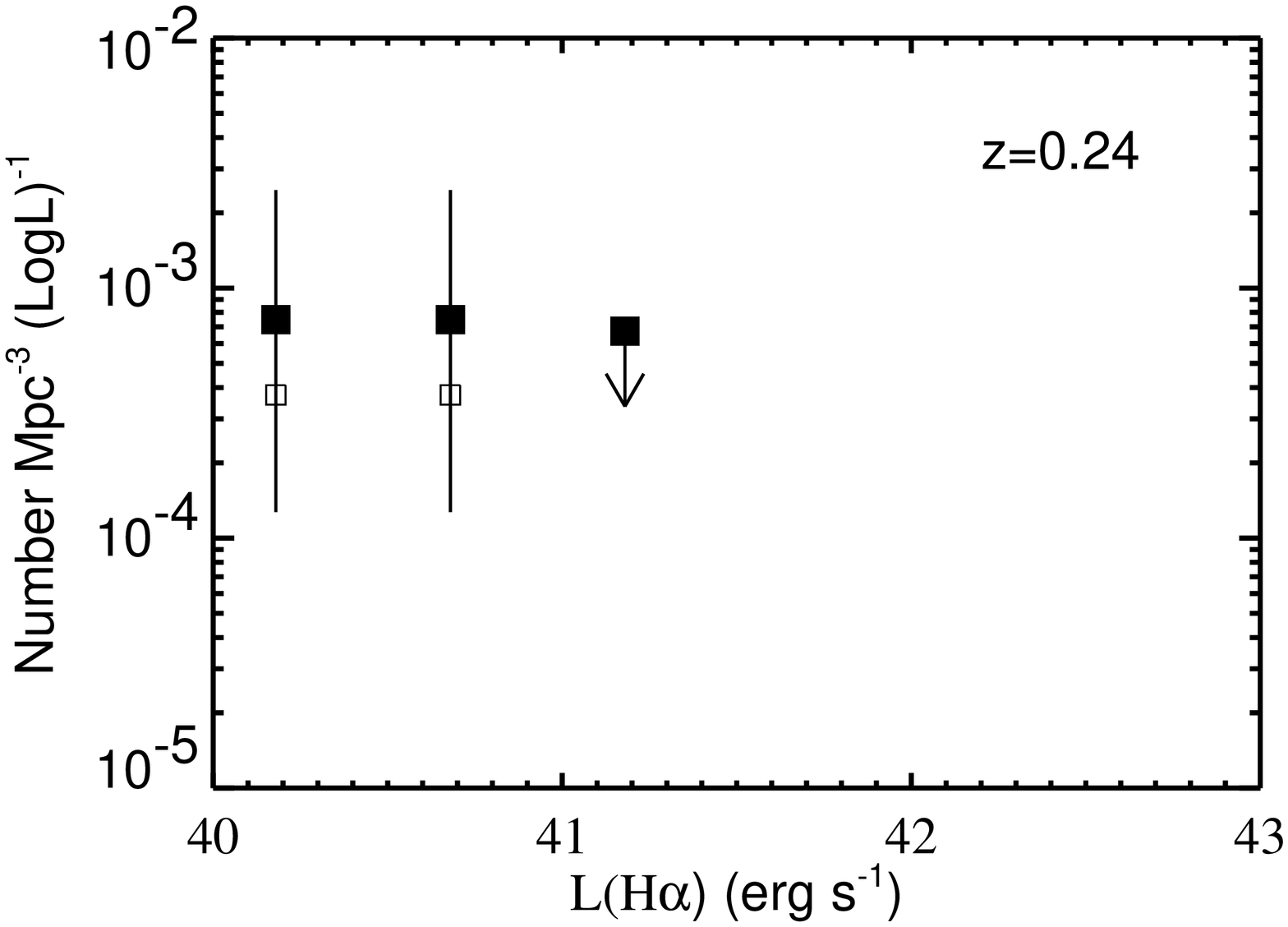}
  \includegraphics[viewport=20 0 460 646,width=2.8in,clip,angle=90,scale=0.85]{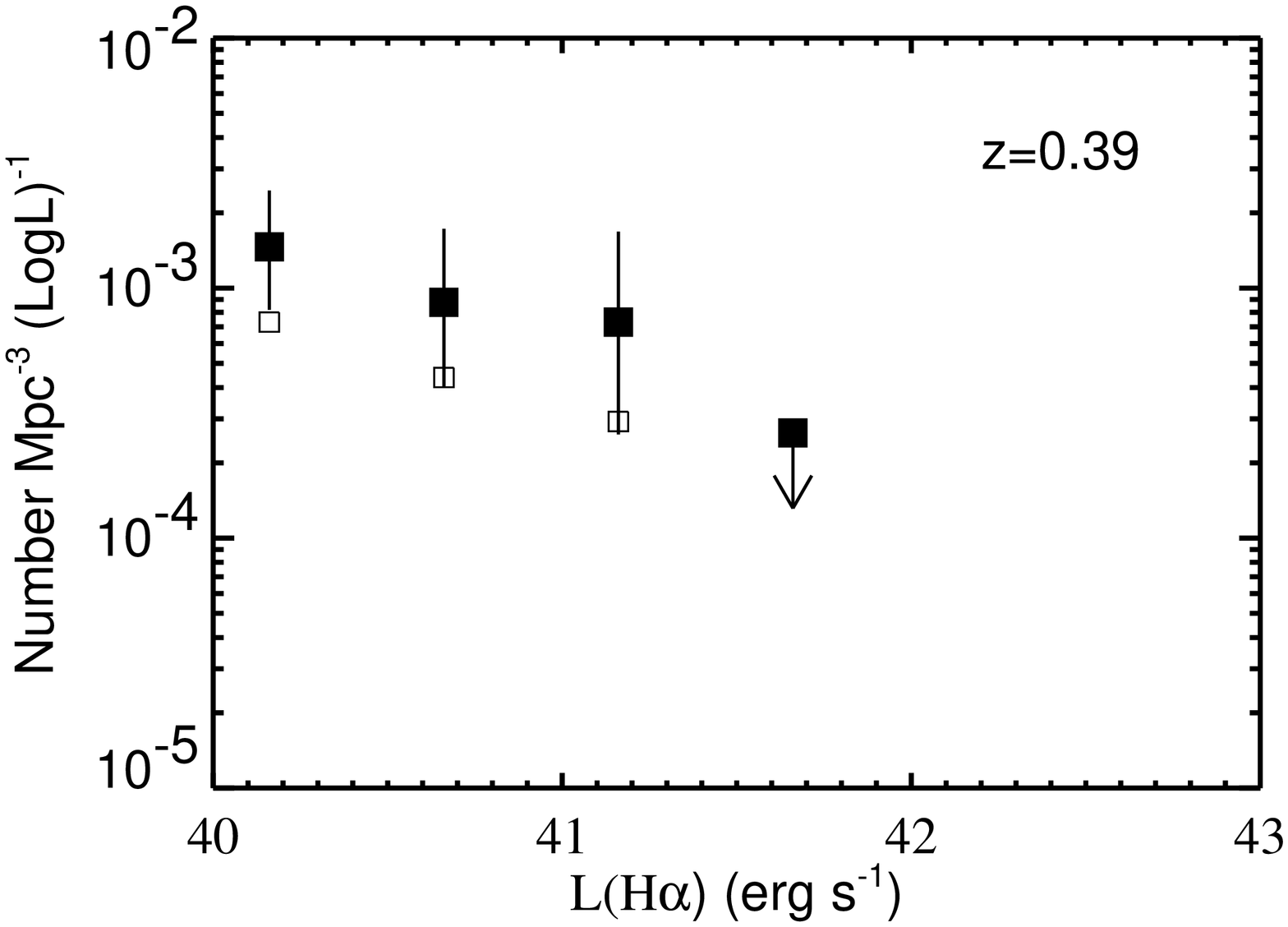}
  \caption{The luminosity function of \protect{\ha}
	  at $z=0.24$ (top panel) and at
	  $z=0.39$ (bottom panel). In each case the open boxes
	  show the luminosity functions determined from the
	  spectroscopic sample alone while the solid boxes show
	  the function corrected for the incompleteness in the
	  spectroscopic identification. The errors are plus and minus
	  1 sigma and at the highest luminosity we show the 1 sigma
	  upper limit.
	  \label{fig6:lum-ha}
  }
  \end{figure}

  \begin{figure}
  \includegraphics[viewport=20 0 456 646,width=2.8in,clip,angle=90,scale=0.85]{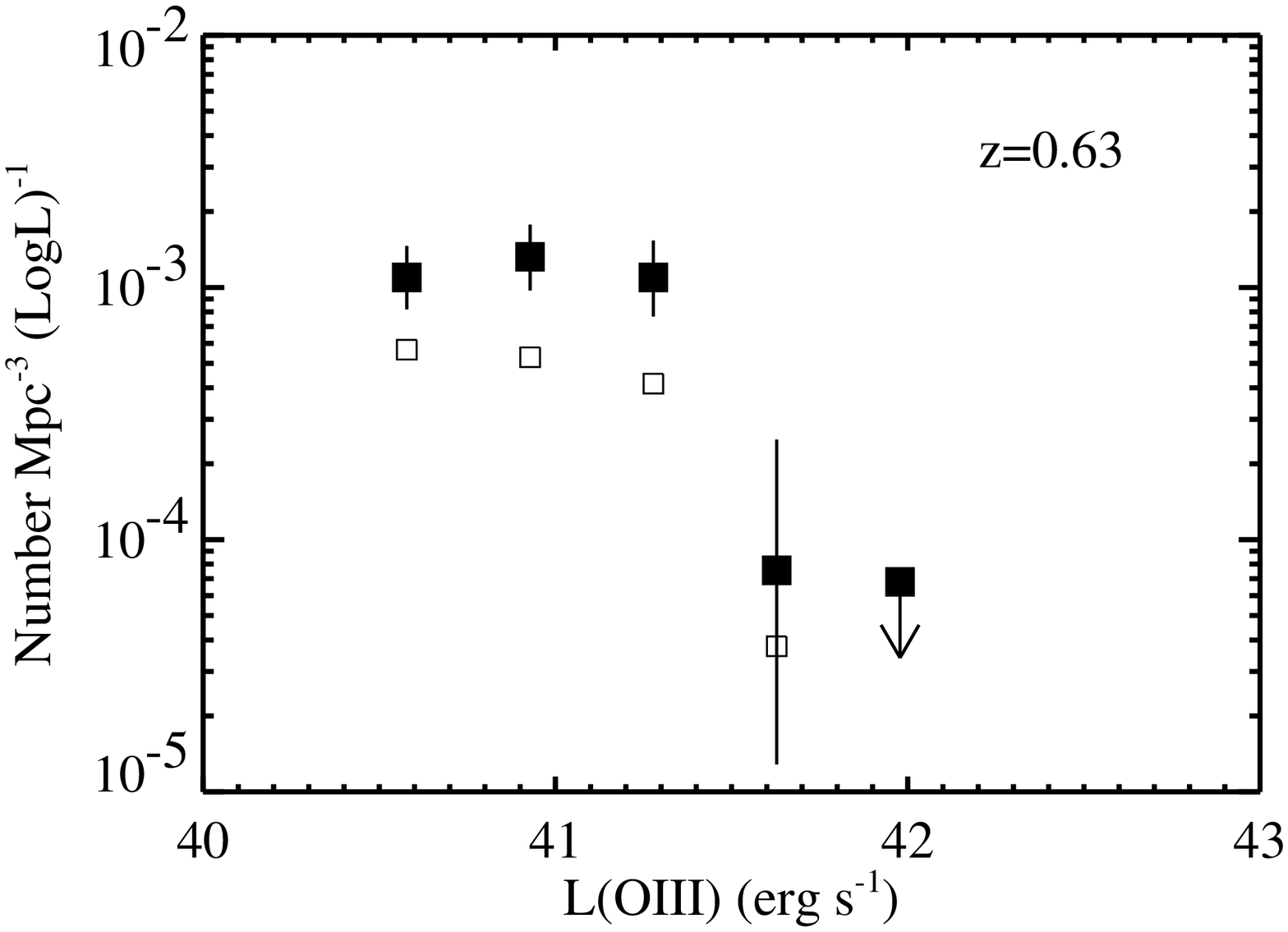}
  \includegraphics[viewport=20 0 456 646,width=2.8in,clip,angle=90,scale=0.85]{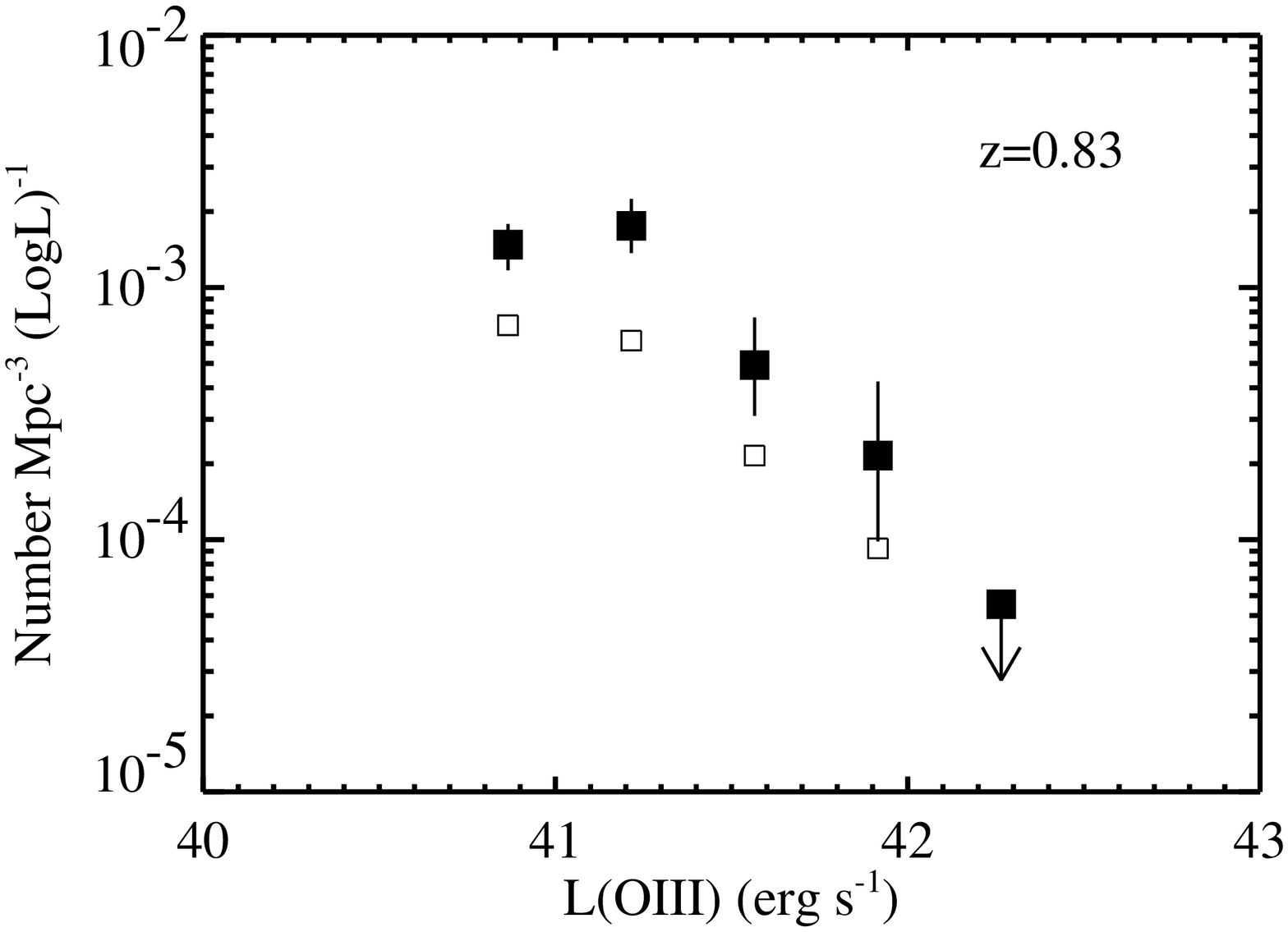}
  \caption{The luminosity function of   \protect{\oiii} $\lambda$5007 emitters
	  at $z=0.63$ (top panel) and at
	  $z=0.83$ (bottom panel). In each case the open boxes
	  show the luminosity functions determined from the
	  spectroscopic sample alone while the solid boxes show
	  the function corrected for the incompleteness in the
	  spectroscopic identification. The errors are plus and minus
	  1 sigma and at the highest luminosity we show the 1 sigma
	  upper limit.
	  \label{fig5:lum-oiii}
  }
  \end{figure}

Because of the high observed frame equivalent widths the primary
fluxes are insensitive to the continuum determination.  However,
they do depend on the filter response at the emission line
wavelength so we first restrict ourselves to redshifts where the
nominal filter response is greater than 70\% of the peak value.
This also has the advantage of providing a uniform selection and
we assume the window function is flat over the defined redshift
range.  Now the volume is simply defined by the selected
redshift range for all objects above the minimum luminosity
which we take as corresponding to a flux of $1.5\times10^{-17}$
erg cm$^{-2}$ s$^{-1}$ (Figure~\ref{fig7:emit-zdistrib}).  The
luminosity function is now obtained by dividing the number of
objects in each luminosity bin by the volume.  The
incompleteness corrected luminosity function is obtained from
the sum of the weights in each luminosity bin divided by the
volume. The 1 sigma errors shown are calculated from the
Poissonian errors based on the number of objects in the bin.
The calculated H$\alpha$  luminosity function  is shown for the
two redshift ranges corresponding to the NB816 and NB912
selections in Figure~\ref{fig6:lum-ha} and the corresponding
\oiii$\lambda$5007 luminosity functions in
Figure~\ref{fig5:lum-oiii}.

\subsection{Star Formation Rates}
\label{secevol}

  \begin{figure}
  \begin{centering}
  \includegraphics[viewport=20 0 446 620,width=2.8in,clip,angle=90,scale=0.8]{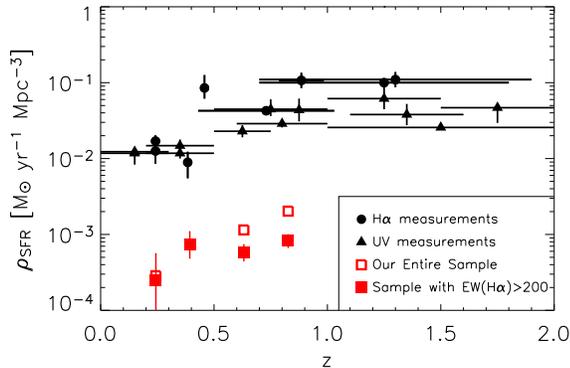}
  \caption{The star formation history inferred from the H$\alpha$ or H$\beta$ 
          luminosity density as a function of redshift. The data from our sample are
          shown in red. The open squares show 
          the total rate from the entire sample while the solid squares show the 
          values for objects with H$\alpha$ rest frame equivalent widths in 
          excess of 200\AA\ or H$\beta$ rest frame equivalent
          widths in excess of 70\AA. The diamonds show the UV star formation
          rates (uncorrected for extinction) from the ground based work of 
          \citet{wilson} and the triangles the Galex results of \citet{wyd05} 
          and \citet{galex}.  H$\alpha$ measurements from the literature as
          summarized in \protect{\citet{ly07}} are shown with filled circles.
          In all cases the errors are $\pm 1 \sigma$.
	  \label{fign:line-ratios2}
  }
  \end{centering}
  \end{figure}

The individual objects have H$\alpha$ luminosities stretching up
to about $3\times10^{41}$ erg s$^{-1}$ where, at the higher
redshifts, we use the H$\beta$ luminosity to infer the H$\alpha$
value assuming there is no reddening.  For a steady formation
this would require a star formation rate of a few solar masses
per year if we adopt the \citet{kennrev} conversion rate for his
Salpeter mass function.

Since the objects are more probably caused by starbursts the
true star formation rate will depend on the evolutionary
history. However, the H$\alpha$ luminosity density should give a
reasonable estimate of the universal star formation density of
the objects provided only that most of the ionizing photons are
absorbed in the galaxies. We first formed the total H$\alpha$ or
H$\beta$ luminosity density of the galaxies by summing over the
incompleteness weighted luminosities in each redshift interval.
We only included detected objects and did not attempt to
extrapolate to lower luminosities but the result are not
particularly sensitive to this because the luminosity functions
are relatively flat at the lower luminosities
(Figures~\ref{fig5:lum-oiii} and \ref{fign:line-ratios2}). We
then converted these to star formation rates with the
\citet{kennrev} conversion.

The results are shown in Figure~\ref{fign:line-ratios2}. We
first plot the rate for the total samples at each redshifts
shown by the open squares. We have shown star formation rates
for UV continuum samples for comparison and the present sample
of strong emitters gives star formation rates which are about
10\% of the UV values over the redshift interval. For comparison,
we also show the star formation rates from H$\alpha$ selected
samples reported in the literature and summarized in \citet{ly07}.
In order to better understand the evolution we have also restricted 
our own sample to objects with rest frame equivalent widths of 
H$\alpha$ in excess of 200\AA\ at low redshifts and H$\beta$ 
equivalent widths in excess of 70\AA\ at the higher redshifts. The 
star formation rates for this sample are shown with the solid 
squares. This provides a more uniform selection with redshift 
and gives a slower increase than the total inferred star formation rate.
For this sample the SFR is evolving roughly as (1+z)$^{3}$
broadly similar to other UV and optically determined formation
rates in this redshift interval. These more restricted objects
comprise about 5\% relative to the UV star formation rates at
the higher redshifts.

\section{Galaxy Metallicities}
\label{secmet}

\subsection{\oiii\ emitters}
\label{o3met}

The spectra are of variable quality and, in order to measure the
metallicities, we need very high signal to noise observations.
It is also important that Balmer lines are well detected since
our flux calibrations are done using the neighboring Balmer
lines (\S4.4).  We therefore restricted ourselves to
\oiii\ emitters whose H$\beta$ fluxes are detected above 15
sigma.  Among 92 \oiii\ emitters in our total spectroscopic
sample, 8 such \oiii\ emitters were chosen in the NB912 sample,
and 10 in the NB816.  These emitters have H$\gamma$ detected
above 4 sigma.  Tables~\ref{tbl-4} and \ref{tbl-5} give the
oxygen line fluxes normalized by their H$\beta$ fluxes for the
NB816 and NB912 selected emitters, respectively.  1$\sigma$
upper limits are listed when the measured flux is below
1$\sigma$.

The most direct method to estimate the gas-phase oxygen
abundance is to use the electron temperature of the HII region.
Higher gas metallicity increases nebular cooling, leading to
lower electron temperatures. Therefore electron temperature is a
good indicator of the gas metallicity.  The electron temperature
can be derived from the ratio of the \oiii$\lambda$4363 auroral
line to \oiii$\lambda$$\lambda$5007,4959. This procedure is
often referred to as the `direct' method or $T_e$ method
\citep[e.g.,][]{seaton,pagel92,pilyugin,izo06c}.  One well-known
problem with the direct method, however, is that
\oiii$\lambda$4363 is generally very weak even in the
low-metallicity galaxies. For higher metallicity systems, the
far-IR lines become the dominant coolant and therefore the
optical auroral line is essentially not detectable.  However,
the majority of our sample exhibit \oiii$\lambda$4363, already
suggesting that these are metal-deficient systems.  To derive
$T_e$\oiii\ and the oxygen abundances, we used the
\citet{pagel92} calibrations with the
$T_e$\oii$-$$T_e$\oiii\ relations derived by \citet{garnett92}.
The results are shown in Table~\ref{tbl-4} (for NB816
selected \oiii\ emitters) and Table~\ref{tbl-5} (for NB912 selected
\oiii\ emitters).  The \citet{izo06c} formula, which was
developed with the latest atomic data and photoionization
models, gives consistent abundances within 0.1 dex.  The
\sii$\lambda\lambda$6717, 6731 lines that are usually used for
the determination of the electron number density, are beyond the
Keck/DEIMOS wavelength coverage for our \oiii\ emitters.
Therefore we assumed n$_e$ = 100 cm$^{-3}$.  However the choice
of electron density has little effect as electron temperature is
insensitive to the electron density; we get the same results
even when we use n$_e$ = 1000 cm$^{-3}$.

The 1$\sigma$ upper and lower limits on $T_e$\oiii\ and the
oxygen abundances are also shown in the tables.  Because the
\oiii$\lambda$4363 flux is weak ($< 4 \sigma$), the range of our
metallicity measurements are quite wide.
However, of 18 \oiii\ emitters, even the upper metallicity
limits on 6 emitters satisfy the definition of XMPGs [12 +
log(O/H) $<$ 7.65; \citealt{kunth}].
All our emitters, except the ones that only have lower limits
on metallicties such as ID31 in Table~\ref{tbl-4}, have very low
metallicities -- even the {\it upper} metallicity limits are
about 0.02 - 0.2 $Z_\sun$.
A few emitters may even have metallicities that are comparable
to the currently known most metal-poor galaxies [I Zw 18 and
SBS0335$-$052W; 12 + log(O/H) $\sim$ 7.1 $-$ 7.2].  However our
current metallicity errors are too large to measure the baseline
metallicity accurately and higher S/N spectra will be necessary
for this purpose.

Our discovery rate of XMPGs appers to be significantly higher
than those of other surveys.
Only 14 new XMPGs have been discovered from the analysis of
$\sim$530,000 galaxy spectra in the SDSS and they are all
located nearby ($z < 0.005$) (SDSS DR3: \citealt{knia03}, SDSS
DR4: \citealt{izo06a}).  At higher redshift, 17 metal-poor ($7.8
< 12$ + log(O/H)$ < 8.3$)
galaxies have been found at $z \sim 0.7$ in the initial phase of
the DEEP2 survey of 3,900 galaxies and the Team Keck Redshift
Survey of 1,536 galaxies \citep{hoy05}.  But none of these
galaxies satisfies the condition of XMPGs.

The present sample may be the first XMPGs at intermediate
redshift ($z \sim 1$) whose oxygen abundances are securely
measured by the direct method.  The narrowband method is very
powerful for finding not only high-redshift ($z >> 5$) galaxies,
but also strong emission-line, extremely metal-deficient
galaxies at intermediate redshifts ($z < 1$).

  \begin{figure}
  \begin{centering}
  \includegraphics[viewport=20 0 510 650,width=3.0in,clip,angle=90,scale=0.8]{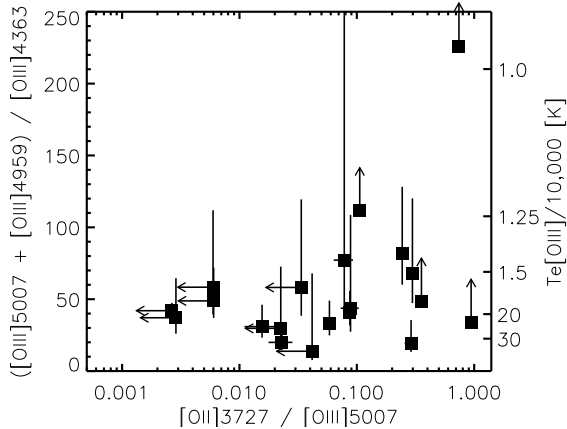}
   \caption{\protect{\oiii}$\lambda$4959$+$$\lambda$5007/\protect{\oiii}$\lambda$4363 versus 
           \protect{\oii}$\lambda$3727/\protect{\oiii}$\lambda$5007 for the \protect{\oiii} and H$\alpha$
            selected emitters in Table~\ref{tbl-4} and \ref{tbl-5}.  The electron temperature of the HII region is also shown.
  \label{fig15:oiii-line-ratios}}
  \end{centering}
  \end{figure}

Figure~\ref{fig15:oiii-line-ratios} shows the electron
temperature sensitive line ratio,
\oiii($\lambda$4959+$\lambda$5007)/\oiii$\lambda$4363 versus
\oii$\lambda$3727/\oiii$\lambda$5007.  If we have an estimate of
the metallicity, as in the present case, we can use the
\oii$\lambda$3727/\oiii$\lambda$5007 ratio to estimate the
ionization parameter $q$. The ionization parameter $q$ is
defined as the number of hydrogen ionizing photons passing
through a unit area per second per unit hydrogen number density
\citep{kewley02}.  For the low metallicity systems with strong
\oiii$\lambda$4363 lines, we can see from
Figure~\ref{fig15:oiii-line-ratios} that \oii$\lambda$3727 is
very weak compared to \oiii$\lambda$5007 with values ranging
downwards from 0.3. Assuming the metallicity is less than 0.2
$Z_\sun$ this would place a lower bound of $q=10^{8}$ cm
s$^{-1}$ on the ionization parameter based on the
\citet{kewley02} model. The higher metallicity objects have
stronger \oii$\lambda$3727/\oiii$\lambda$5007 which, while in
part due to the metallicity, also requires these objects to have
lower ionization parameters suggesting we are seeing an
evolutionary sequence.

  \subsection{H$\alpha$ emitters}
  \label{Hamet}
   Among 13 H$\alpha$ emitters in our spectroscopic sample, only
   3 NB912 selected emitters have H$\beta$ fluxes adequate ($>
   15\sigma$) for the purpose of metallicity measurements.
   Their $T_e$\oiii\ and oxygen abundances were measured
   using the direct method described above, and are shown in the
   Table~\ref{tbl-5} together with the data for the \oiii\ emitters.  The
   \oii$\lambda$3727 line of ID266 is outside the Keck/DEIMOS
   wavelength coverage. In order to derive an upper-limit on the
   metallicity, we assumed \oii$\lambda$3727/\oiii$\lambda$5007
   = 1.0, which is the maximum value in our sample (see,
   Fig.~\ref{fig15:oiii-line-ratios}).  All our H$\alpha$
   emitters are metal poor ($Z_{\rm upper}$ $<$ 0.45 $Z_\sun$),
   but none of them are XMPGs.

\subsection{Composite Spectrum}
\label{secompos}

As can be seen in Figure~\ref{fig15:oiii-line-ratios} the
objects with low \oii$\lambda$3727/\oiii$\lambda$5007 have
relatively uniform values of
(\oiii$\lambda$5007+$\lambda$4959)/\oiii$\lambda$4363 and
similar metallicities. Given the relatively low signal to noise
of the individual spectra it therefore seems of interest to form
a composite spectrum. Such a spectrum will have weightings on
the lines which depend on the individual ionization parameters
and metallicity but will give a rough estimate of the average
metallicity and temperature of the sample.

\begin{figure*}
  \begin{centering}
  \includegraphics[viewport=20 0 446 616,width=4.8in,clip,angle=90,scale=0.8]{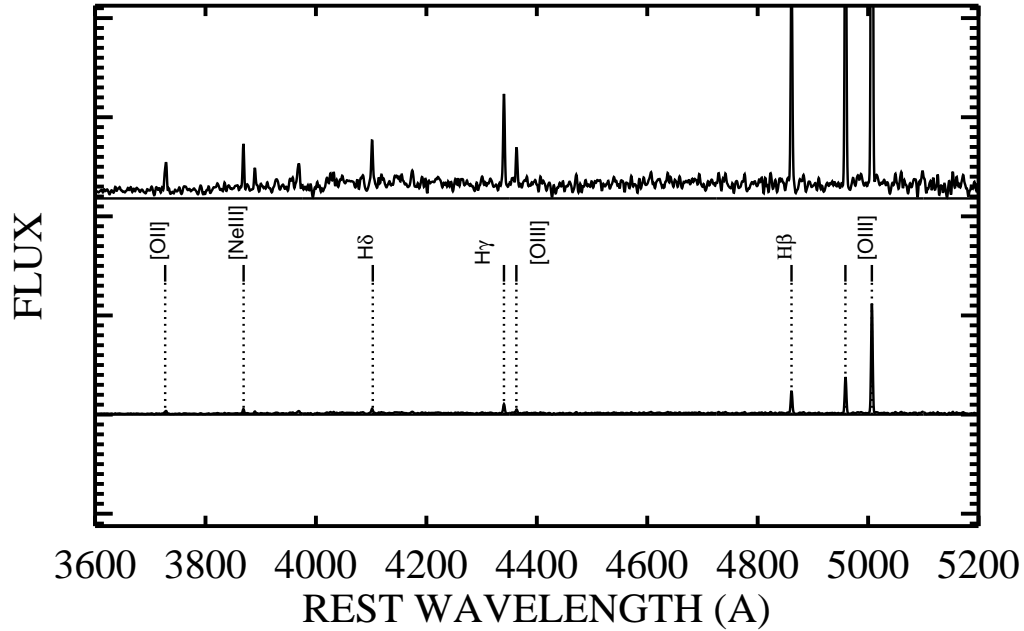}
  \caption{Composite spectrum of the 8 emitters with \protect{\oii}$\lambda$3727/\protect{\oiii}$\lambda$5007 less than 0.1.
  The eight spectra have simply been summed without weighting. The lower panel shows the
  stronger lines and the upper the continuum and the weaker lines. The stronger emission
  lines are labelled and marked with the vertical dotted lines.
  \label{fig16:compos_spec}}
  \end{centering}
  \end{figure*}

In Figure~\ref{fig16:compos_spec} we show the composite spectrum
of the 8 objects with \oii$\lambda$3727/\oiii$\lambda$5007 less
than 0.1. The \oiii$\lambda$4363 is now strongly detected with a
value of $16.7\pm2.1$ or eight sigma.  The mean temperature is
$19,500\pm1,500$ K and the average abundance
12+log(O/H)=$7.5\pm0.1$ and the mean rest frame equivalent width
of H$\beta$ is 57\AA . The results are similar to the values
obtained by averaging the individual analyses of the eight
objects.

\section{Morphologies}
\label{secmorph}

The morphology of the USELs may give us a clue to the mechanism
of their high star formation activity (SFR $\sim$ a few
$M_\sun$/yr) and star formation history; what has triggered the
star formation $-$ mergers, gas infall, or galactic winds?  High
resolution HST/ACS images are available for GOODS-North
(GOODS-N) region \citep{gia04a} which is one of our survey
fields.  There are 17 NB816 selected USELs in the GOODS-N, and
16 in the NB912 sample.  Figures~\ref{figy:hst_816} and \ref{figz:hst_912} show thumbnails of
the NB816 and NB912 selected USELs in the GOODS-N field (respectively)
with each thumbnail $12\farcs5$  on a side.  The white
dashes point to the largest galaxy.  We used continuum broadband
images to show underlying stellar populations:  HST/ACS B, V,
$z'$-band images were used for NB816 emitters, and B, V, I-band
for NB912 emitters. High-redshift Ly$\alpha$ emitters ($z >> 5$)
are very red because of the continuum absorption below
Ly$\alpha$ emission caused by neutral hydrogen in the
intergalactic medium.  We do not have spectra for most of the
USELs in the GOODS-N field yet, and none of the USELs in GOODS-N
have metallicity measurements either due to lack of spectra or
low spectral S/N.  But we can qualitatively argue that the USELs
at intermediate redshift ($z < 1$) exhibit widespread
morphologies from relatively compact high surface brightness
objects to very diffuse low surface brightness ones.

\section{Discussion}
\label{secdisc}

  \begin{figure*}
  \begin{centering}
  \figurenum{19}
   \includegraphics[viewport=20 0 500 610,width=3.6in,clip,angle=90,scale=1.0]{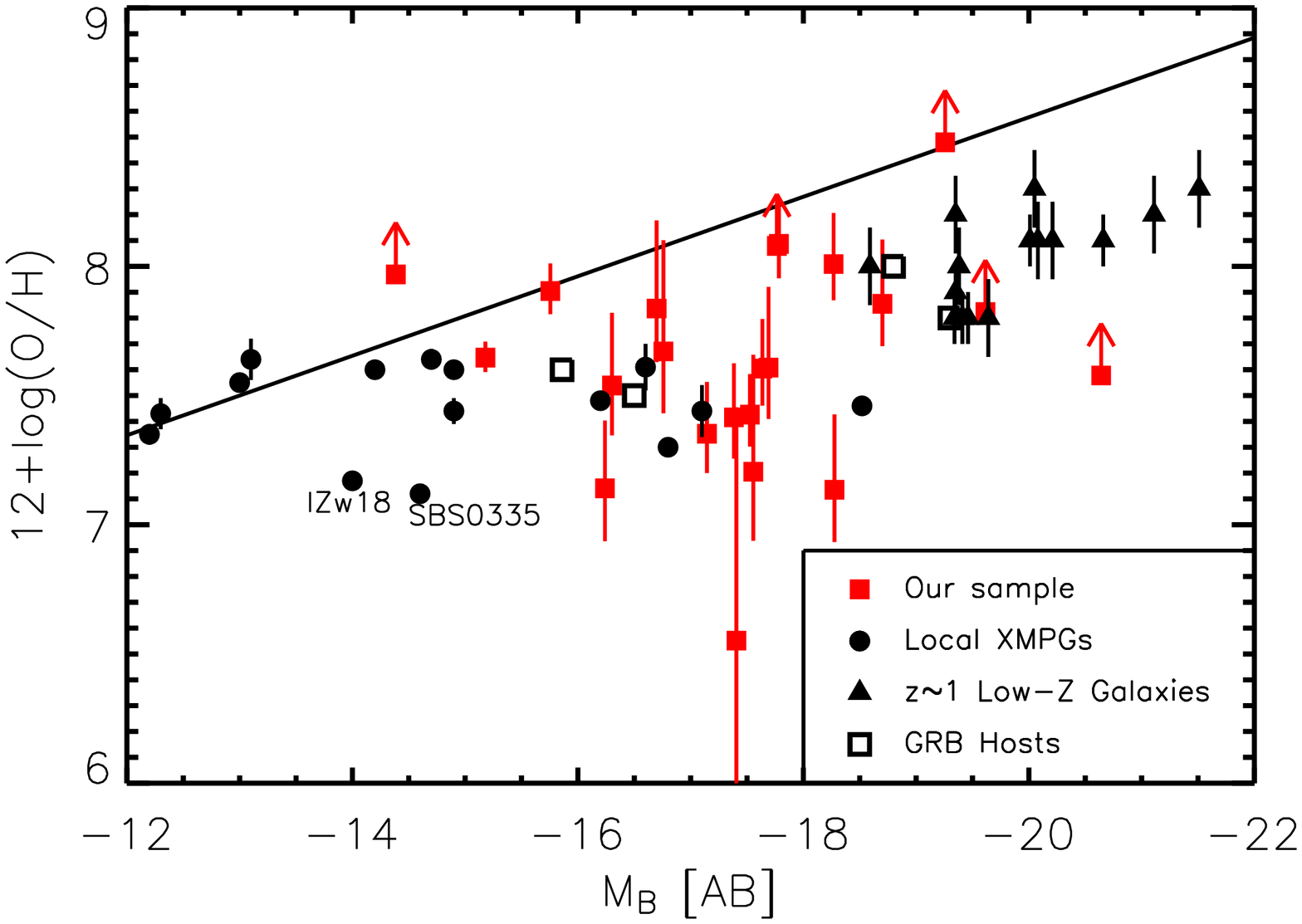}
   \caption{The oxygen abundance versus the absolute rest frame $B$ magnitude
            for the \oiii\ selected samples ({\it red squares}).  
            One sigma errors are shown for the oxygen abundances and one sigma lower 
            limits are shown with upward pointing arrows.
            The solid line shows the \citep{skil89} relation 
            for the nearby dwarf irregulars. As with the local XMPGs 
            ({\it filled circles}, \citealt{knia03,kewley07} 
            and GRB hosts (open squares, \citealt{stanek06,kewley07}, the 
            present galaxies are much more luminous at a given metallicity than the local irregulars.
            Metal-poor luminous galaxies (but not XMPGs) at $z\sim1$ from \citealt{hoy05}. are shown as triangles.
            A few of our emitters may have oxygen abundances comparable to the most
            metal-deficient galaxis, I Zw 18 (12 $+$ log O/H = $7.17 \pm 0.01$, \citealt{thuan05}) 
            and SBS 0335-052W (12 $+$ log O/H = $7.12 \pm 0.03$, \citealt{izo05}).
	\label{figm:metal_mags}
}
\end{centering}
\end{figure*}

The present emitters differ from the local
dwarf HII galaxies in a large number of ways though they
appear much more similar to the XMPGs found in the
SDSS samples. They are much more
luminous, have large \oiii/\oii\ ratios, and they are a relatively
high fraction (about 10\% by number from
Figure~\ref{fig5:lum-oiii}) of other galaxy populations at these redshifts.
Taken together this suggests that we are seeing much more
massive galaxies in the early stages of formation and,
since we need these to have relatively long lifetimes
in order to understand their frequency, that we are seeing
objects undergoing continuous star formation rather than
single starbursts. For the case of constant star formation
with a standard Salpeter IMF a forming galaxy can have
equivalent widths above 30\AA\ for $10^{9}$ yr \citep{star99}
which would allows us to understand the observed
number density of strong emitters relative to the total
galaxy population.

In this type of model we would expect the metallicity to grow
with time and that higher metallicity galaxies would have higher
continuum magnitudes and lower equivalent widths in H$\beta$. We
plot the absolute rest frame $B$ magnitudes versus the Oxygen
abundance in Figure~\ref{figm:metal_mags}.  As with the case for
the local XMPGs found in the SDSS ({\it filled circles},
\citealt{knia03,kewley07}) and the metal-poor galaxies ($7.8 <$
12 + log O/H $< 8.3$) at $z \sim 1$ ({\it triangles},
\citealt{hoy05}),
the present emitters ({\it red squares}) are much more luminous
at a given metallicity than is found for the local dIrrs ({\it
solid line}, \citealt{skil89}).  Furthermore there does indeed
seem to be a trend to higher continuum luminosities at higher
metallicity consistent with ongoing star formation raising the
luminosity.  Recently \citet{kewley07} reported the similarity
between XMPGs and long duration GRB hosts ; they share similar
SFRs, extinction levels, and both lie in a similar region of the
luminosity-metallicity diagram. Our sample metal-deficient
galaxies, which also lie in the region of GRB hosts, may be
additional support of the connection between XMPGs and GRB
hosts.

Of the six galaxies with continuum magnitudes brighter
than -18 all but one have metallicities or lower limits
which would place them near or above 12+log(O/H)=8
while the lower luminosity galaxies primarily have
12+log(O/H) in the range $7.1-7.8$. If we assumed that the 
metallicity were a simple linear function of the age then the
more luminous galaxies would be several times older
than the less luminous ones which is not quite enough
to account for the luminosity increase (see e.g.
\citealt{star99}) suggesting that the enrichment 
process may be more complex. However, the accuracy
of our current metallicity measurements may be inadequate
for measuring the lowest metallicities in the sample and
we could be underestimating the amount of metallicity evolution.

  \begin{figure}
  \begin{centering}
  \figurenum{20}
  \includegraphics[viewport=20 0 500 660,width=3.0in,clip,angle=90,scale=0.8]{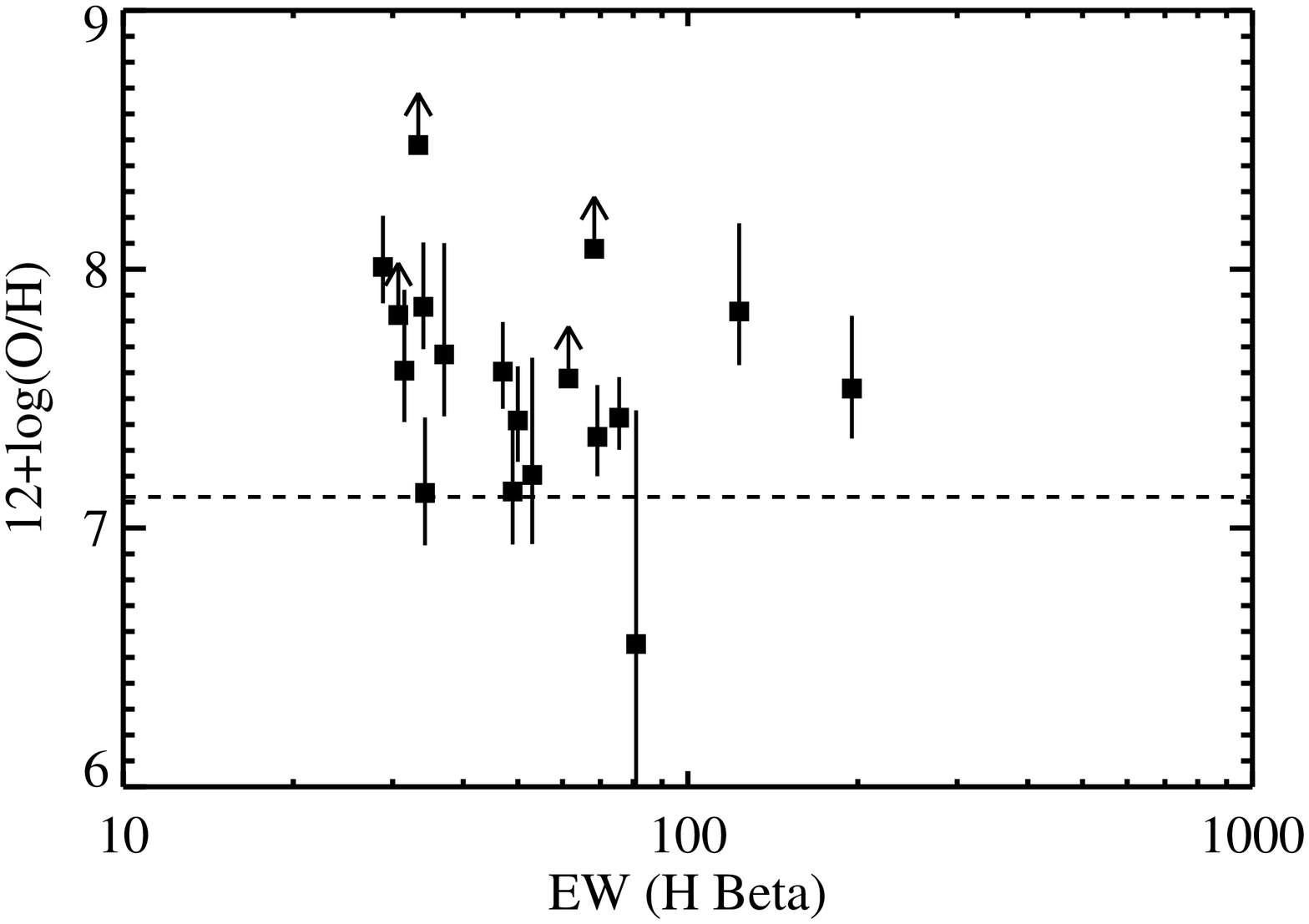}
  \caption{The oxygen abundance versus the rest frame H$\beta$ equivalent width 
           for the \protect{\oiii} selected samples. 
           One sigma errors are shown for the
           oxygen abundances and one sigma lower limits are shown with upward pointing arrows.
           The dotted line shows the metallicity of 1Zw-18.
	   \label{figx:metal_ew}
  }
  \end{centering}
  \end{figure}

The relation between the metallicity and the H$\beta$ equivalent
width is shown in Figure~\ref{figx:metal_ew}. There clearly is a
large scatter in metallicity at all equivalent widths suggesting
that the star formation may be episodic with periods in which
bursts of star formation enhances the H$\beta$ equivalent widths
in objects where previous star formation has raised the
metallicity.

With better spectra and more accurate metal estimates we should
be able to refine these tests and also determine whether the number
density of objects versus metallicity is consistent with that expected
in a simple growth model. Perhaps even more importantly as larger
spectroscopic samples are obtained we should be able to determine
if there is a floor on the metallicity corresponding to the
enrichment in the intergalactic gas. Within the errors we have
yet to find an object with lower metallicity than the lowest
metallicity local galaxies but this could easily change as the
observations are improved.

\section{Summary}
\label{secsum}

We have described the results of spectrscopic observations of a
narrowband selected sample of extreme emission line objects. The
results show that such objects are common in the $z=0-1$
redshift interval and produce about 5-10\% of the star formation
seen in ultraviolet or emission line measurements at these
redshifts. A very large fraction of the strong emitters are
detected in the \oiii$\lambda$4363 line and oxygen abundances
can be measured using the direct method.  The abundance
primarily lie in the 12+log(O/H) range of 7$-$8 characteristic
of XMPGs.

The results suggest that we are seeing early chemical enrichment
of startup galaxies at these redshifts which are forming in
relatively chemically regions. As we develop larger samples of
these objects and improve the accuracy of their abundance
estimates we should be able to test this model and to determine
if there is a floor on the metallicity of the galaxies.

\acknowledgements
We are indebted to the staff of the Subaru and Keck
observatories for their excellent assistance with the
observations. We acknowledge Subaru/SuprimeCam support
astronomer, Hisanori Furusawa, for his help over several years
with the observations.  Y. Kakazu acknowledges invaluable advice
from Lisa J. Kewley and Roberto Terlevich on metallicity
measurements. This work was supported in part by NSF grants
AST04-07374 (LLC) and AST06-87850 (EMH), and Spitzer grant JPL
1289080 (EMH).


\newpage
\begin{deluxetable}{rccccccccr}
\tablecolumns{10}
\tablecaption{NB816 Emission-Line Sample}
\tablewidth{0pt}
\tablehead{
\colhead{No.}        & \colhead{RA(2000)}    &
\colhead{Dec(2000)}     & \colhead{$N(AB)$}     &
\colhead{$Z(AB)$}       & \colhead{$I$}         &
\colhead{$R$}           & \colhead{$V$}         &
\colhead{$B$}           & \colhead{$z$}         }

\scriptsize
\startdata
\phn{1}  &    40.115555  &    $-$1.694722  &  23.92  &  25.28  &  24.84  &  25.44  & $-$99.00  & $-$99.00  &  $-$1.0000 \\
 \phn{2}  &    40.116665  &    $-$1.617361  &  24.26  &  25.88  &  25.52  &  25.74  & $-$99.00  & $-$99.00  &  $-$1.0000 \\
 \phn{3}  &    40.138332  &    $-$1.405639  &  23.49  &  24.94  &  24.70  &  25.22  & $-$99.00  & $-$99.00  &   0.6343 \\
 \phn{4}  &    40.174721  &    $-$1.704750  &  24.27  &  25.46  &  25.35  &  25.82  & $-$99.00  & $-$99.00  &   0.6355 \\
 \phn{5}  &    40.183056  &    $-$1.495417  &  24.58  &  25.28  &  25.85  &  26.86  & $-$99.00  & $-$99.00  &   5.6886 \\
 \phn{6}  &    40.216946  &    $-$1.494805  &  24.80  &  26.08  &  26.31  &  26.14  & $-$99.00  & $-$99.00  &   0.2416 \\
 \phn{7}  &    40.250832  &    $-$1.744639  &  23.51  &  24.35  &  24.33  &  24.54  & $-$99.00  & $-$99.00  &  $-$1.0000 \\
 \phn{8}  &    40.276112  &    $-$1.518139  &  24.72  &  24.78  &  25.53  &  25.75  & $-$99.00  & $-$99.00  &  $-$1.0000 \\
  \phn{9}  &    40.276390  &    $-$1.623250  &  24.55  &  25.36  &  25.36  &  25.76  & $-$99.00  & $-$99.00  &  $-$1.0000 \\
  10  &    40.284168  &    $-$1.453583  &  21.82  &  23.20  &  23.02  &  22.56  & $-$99.00  & $-$99.00  &   0.2480 \\
  11  &    40.298889  &    $-$1.447389  &  24.32  &  25.95  &  25.42  &  25.71  & $-$99.00  & $-$99.00  &  $-$1.0000 \\
  12  &    40.304165  &    $-$1.391694  &  20.88  &  22.35  &  22.06  &  22.32  & $-$99.00  & $-$99.00  &  $-$1.0000 \\
  13  &    40.306946  &    $-$1.638500  &  24.91  &  25.67  &  25.72  &  26.02  & $-$99.00  & $-$99.00  &  $-$1.0000 \\
  14  &    40.311111  &    $-$1.535111  &  24.08  &  25.93  &  25.49  &  25.71  & $-$99.00  & $-$99.00  &   0.6240 \\
  15  &    40.318333  &    $-$1.548222  &  24.60  &  25.03  &  25.50  &  25.96  & $-$99.00  & $-$99.00  &  $-$1.0000 \\
  16  &    40.318890  &    $-$1.430889  &  23.37  &  23.77  &  24.32  &  24.42  & $-$99.00  & $-$99.00  &   1.1804 \\
  17  &    40.319168  &    $-$1.446333  &  23.60  &  23.85  &  24.43  &  24.56  & $-$99.00  & $-$99.00  &   1.1873 \\
  18  &    40.320835  &    $-$1.778028  &  20.70  &  21.57  &  21.60  &  21.86  & $-$99.00  & $-$99.00  &  $-$1.0000 \\
  19  &    40.324165  &    $-$1.409972  &  24.29  &  24.38  &  25.18  &  25.45  & $-$99.00  & $-$99.00  &  $-$1.0000 \\
  20  &    40.326111  &    $-$1.709805  &  23.24  &  24.75  &  24.62  &  24.96  & $-$99.00  & $-$99.00  &  $-$1.0000 \\
  21  &    40.336388  &    $-$1.570194  &  24.26  &  24.73  &  25.06  &  25.33  & $-$99.00  & $-$99.00  &  $-$1.0000 \\
  22  &    40.337223  &    $-$1.388194  &  24.81  &  27.37  &  26.71  &  26.77  & $-$99.00  & $-$99.00  &  $-$1.0000 \\
  23  &    40.337502  &    $-$1.658306  &  24.89  &  25.20  &  25.76  &  26.05  & $-$99.00  & $-$99.00  &  $-$1.0000 \\
  24  &    40.340279  &    $-$1.689472  &  24.55  &  26.37  &  26.07  &  26.51  & $-$99.00  & $-$99.00  &  $-$1.0000 \\
  25  &    40.340557  &    $-$1.551889  &  24.99  &  25.70  &  25.93  &  26.35  & $-$99.00  & $-$99.00  &  $-$1.0000 \\
  26  &    40.340832  &    $-$1.371222  &  22.42  &  23.43  &  23.29  &  23.16  & $-$99.00  & $-$99.00  &  $-$1.0000 \\
  27  &    40.340832  &    $-$1.516250  &  24.89  &  25.06  &  25.69  &  25.78  & $-$99.00  & $-$99.00  &  $-$1.0000 \\
  28  &    40.341110  &    $-$1.493500  &  23.37  &  24.62  &  24.55  &  24.23  & $-$99.00  & $-$99.00  &  $-$1.0000 \\
  29  &    40.341389  &    $-$1.484139  &  24.48  &  25.46  &  25.49  &  25.73  & $-$99.00  & $-$99.00  &  $-$1.0000 \\
  30  &    40.342777  &    $-$1.599528  &  24.65  &  25.31  &  25.50  &  25.85  & $-$99.00  & $-$99.00  &  $-$1.0000 \\
  31  &    40.343056  &    $-$1.438833  &  23.17  &  24.60  &  24.54  &  24.54  & $-$99.00  & $-$99.00  &   0.6226 \\
  32  &    40.347500  &    $-$1.403833  &  24.86  &  27.45  &  26.16  &  26.35  & $-$99.00  & $-$99.00  &   0.6324 \\
  33  &    40.349724  &    $-$1.598472  &  23.11  &  23.87  &  24.11  &  24.46  & $-$99.00  & $-$99.00  &   1.1956 \\
  34  &    40.356388  &    $-$1.515722  &  24.94  &  26.19  &  26.18  &  26.42  & $-$99.00  & $-$99.00  &  $-$5.0000 \\
  35  &    40.372223  &    $-$1.390361  &  23.85  &  24.68  &  24.72  &  24.62  & $-$99.00  & $-$99.00  &  $-$1.0000 \\
  36  &    40.373611  &    $-$1.722528  &  24.93  &  25.85  &  25.74  &  25.88  & $-$99.00  & $-$99.00  &  $-$1.0000 \\
  37  &    40.377777  &    $-$1.701889  &  23.45  &  23.82  &  24.26  &  24.70  & $-$99.00  & $-$99.00  &  $-$1.0000 \\
  38  &    40.388054  &    $-$1.697361  &  23.96  &  24.32  &  24.79  &  24.90  & $-$99.00  & $-$99.00  &  $-$1.0000 \\
  39  &    40.388889  &    $-$1.573361  &  24.79  &  25.13  &  25.65  &  26.07  & $-$99.00  & $-$99.00  &  $-$1.0000 \\
  40  &    40.394444  &    $-$1.521389  &  22.73  &  23.69  &  23.84  &  24.06  & $-$99.00  & $-$99.00  &   0.6292 \\
\enddata
\tablecomments{Magnitudes are measured in 3$''$ diameter apertures.
An entry of `$-$99' indicates that no excess flux was measured. $-1.0000$ in the redshift
column means no spectroscopic data were obtained for the object. This is a sample table showing the first entries of the electronic version of the table that will accompany the published paper.}
\label{tbl-2}
\end{deluxetable}

\newpage
\begin{deluxetable}{cccccccccr}
\renewcommand\baselinestretch{1.0}
\tablewidth{0pt}
\tablecaption{NB912 Emission-Line Sample}
\tabletypesize{\footnotesize}
\tablehead{\\ No.\ & R.A. (J2000.0) & Decl. (J2000.0) & $N(AB)$ & $z^\prime(AB)$ & 
 $I$ & $R$ & $V$ & $B$ & $z_{spec}$ \\ 
(1) & (2) & (3) & (4) & (5) & (6) & (7) & (8) & (9) & (10) }
\startdata
 \phn{1}  &    40.131668  &    $-$1.408361  &  23.86  &  25.08  &  25.89  &  25.97
& $-$99.00  & $-$99.00  &   0.8371 \\
\phn{2}  &    40.133888  &    $-$1.575222  &  24.76  &  25.92  &  26.13  &  25.84
& $-$99.00  & $-$99.00  &   1.4498 \\
\phn{3}  &    40.148056  &    $-$1.593555  &  23.21  &  24.62  &  25.53  &  26.11
& $-$99.00  & $-$99.00  &   0.8207 \\
\phn{4}  &    40.148335  &    $-$1.725417  &  24.76  &  25.97  &  26.98  &  26.39
& $-$99.00  & $-$99.00  &   0.8111 \\
\phn{5}  &    40.150833  &    $-$1.737556  &  23.31  &  24.35  &  24.88  &  25.32
& $-$99.00  & $-$99.00  &   0.8269 \\
\phn{6}  &    40.153332  &    $-$1.536833  &  23.64  &  25.00  &  26.18  &  26.87
& $-$99.00  & $-$99.00  &   0.8301 \\
\phn{7}  &    40.156387  &    $-$1.765833  &  21.93  &  23.05  &  23.51  &  23.74
& $-$99.00  & $-$99.00  &  $-$1.0000 \\
\phn{8}  &    40.165833  &    $-$1.580056  &  24.94  &  26.20  &  25.92  &  25.84
& $-$99.00  & $-$99.00  &   1.4482 \\
\phn{9}  &    40.183334  &    $-$1.389583  &  21.77  &  22.79  &  23.33  &  23.52
& $-$99.00  & $-$99.00  &   0.8325 \\
  10  &    40.184444  &    $-$1.596444  &  23.09  &  24.50  &  25.30  &  25.55
& $-$99.00  & $-$99.00  &   0.8293 \\
  11  &    40.193611  &    $-$1.690083  &  24.32  &  25.40  &  25.83  &  25.69
& $-$99.00  & $-$99.00  &   0.8266 \\
  12  &    40.194168  &    $-$1.373722  &  23.93  &  24.93  &  25.21  &  25.35
& $-$99.00  & $-$99.00  &  $-$1.0000 \\
  13  &    40.194721  &    $-$1.373917  &  24.87  &  25.90  &  26.30  &  26.47
& $-$99.00  & $-$99.00  &  $-$1.0000 \\
  14  &    40.196667  &    $-$1.378333  &  24.07  &  25.42  &  26.05  &  26.18
& $-$99.00  & $-$99.00  &  $-$1.0000 \\
  15  &    40.202221  &    $-$1.584472  &  24.22  &  25.30  &  25.81  &  26.19
& $-$99.00  & $-$99.00  &   0.8289 \\
  16  &    40.203335  &    $-$1.471861  &  24.77  &  26.20  &  25.78  &  25.54
& $-$99.00  & $-$99.00  &   0.3965 \\
  17  &    40.214722  &    $-$1.519917  &  23.14  &  24.40  &  25.38  &  25.94
& $-$99.00  & $-$99.00  &   0.8288 \\
  18  &    40.220276  &    $-$1.753778  &  24.34  &  25.41  &  26.40  &  26.29
& $-$99.00  & $-$99.00  &  $-$1.0000 \\
  19  &    40.220833  &    $-$1.388556  &  23.24  &  24.36  &  25.13  &  24.99
& $-$99.00  & $-$99.00  &  $-$1.0000 \\
  20  &    40.226944  &    $-$1.542111  &  23.02  &  24.47  &  25.77  &  25.48
& $-$99.00  & $-$99.00  &   0.8208 \\
  21  &    40.229168  &    $-$1.720889  &  24.99  &  27.85  &  27.11  &{\nodata}
& $-$99.00  & $-$99.00  &   6.4800 \\
  22  &    40.229443  &    $-$1.376472  &  23.75  &  24.98  &  25.72  &  24.66
& $-$99.00  & $-$99.00  &  $-$1.0000 \\
  23  &    40.245834  &    $-$1.578972  &  24.61  &  25.82  &  27.22  &  26.91
& $-$99.00  & $-$99.00  &   0.8285 \\
  24  &    40.280556  &    $-$1.421056  &  24.82  &  25.86  &  26.13  &  25.91
& $-$99.00  & $-$99.00  &  $-$1.0000 \\
  25  &    40.290833  &    $-$1.746361  &  23.89  &  25.16  &  25.81  &  25.03
& $-$99.00  & $-$99.00  &   0.0000 \\
  26  &    40.323891  &    $-$1.697667  &  24.73  &  25.80  &  26.47  &  25.71
& $-$99.00  & $-$99.00  &  $-$1.0000 \\
  27  &    40.330833  &    $-$1.612389  &  23.03  &  24.68  &  26.06  &  25.47
& $-$99.00  & $-$99.00  &   0.0000 \\
  28  &    40.339722  &    $-$1.395361  &  23.87  &  25.54  &  26.98  &  25.45
& $-$99.00  & $-$99.00  &   0.3889 \\
  29  &    40.346668  &    $-$1.448305  &  23.93  &  24.98  &  25.37  &  25.48
& $-$99.00  & $-$99.00  &   0.8274 \\
  30  &    40.382500  &    $-$1.554056  &  23.52  &  24.87  &  25.60  &  25.08
& $-$99.00  & $-$99.00  &   0.3930 \\
  31  &    40.393055  &    $-$1.713694  &  24.94  &  26.83  &  26.70  &  26.50
& $-$99.00  & $-$99.00  &  $-$1.0000 \\
  32  &    40.398888  &    $-$1.466417  &  23.87  &  24.91  &  24.83  &  24.64
& $-$99.00  & $-$99.00  &   1.4590 \\
  33  &    40.403889  &    $-$1.530167  &  24.88  &  26.08  &  27.44  &  27.30
& $-$99.00  & $-$99.00  &   0.8223 \\
  34  &    40.409443  &    $-$1.369222  &  21.52  &  23.44  &  24.47  &  24.04
& $-$99.00  & $-$99.00  &  $-$1.0000 \\
  35  &    40.411667  &    $-$1.691417  &  24.41  &  25.50  &  25.68  &  25.55
& $-$99.00  & $-$99.00  &  $-$1.0000 \\
  36  &    40.424999  &    $-$1.454028  &  21.91  &  22.93  &  23.42  &  23.41
& $-$99.00  & $-$99.00  &  $-$1.0000 \\
  37  &    40.430000  &    $-$1.501111  &  23.13  &  24.21  &  24.76  &  24.92
& $-$99.00  & $-$99.00  &   0.8267 \\
  38  &    40.446110  &    $-$1.676139  &  24.91  &  26.09  &  26.37  &  26.04
& $-$99.00  & $-$99.00  &  $-$1.0000 \\
  39  &    40.478889  &    $-$1.534278  &  24.87  &  26.06  &  25.79  &  25.56
& $-$99.00  & $-$99.00  &   0.3861 \\
  40  &    40.506111  &    $-$1.755111  &  22.73  &  24.27  &  25.24  &  25.19
& $-$99.00  & $-$99.00  &  $-$1.0000 \\
  41  &    40.511665  &    $-$1.596944  &  24.73  &  25.96  &  25.55  &  25.67
& $-$99.00  & $-$99.00  &  $-$1.0000 \\
  42  &    40.518055  &    $-$1.666139  &  22.47  &  23.73  &  24.37  &  24.45
& $-$99.00  & $-$99.00  &  $-$1.0000 \\
\enddata
\tablecomments{Magnitudes are measured in 3$''$ diameter apertures.
An entry of `$-$99' indicates that no excess flux was measured. $-1.0000$ in the redshift
column means no spectroscopic data were obtained for the object. This is a sample table showing the first entries of the electronic version of the table that will accompany the published paper.}
\label{tbl-3}
\end{deluxetable}

\newpage


\scriptsize

\begin{deluxetable}{ccccccc}
\renewcommand\baselinestretch{1.0}

\tablecaption{Line fluxes and Oxygen Abundance for L816 selected emitters\label{tbl-4}}
\scriptsize
\tablewidth{0pt}

\tablehead{
\colhead{Object}           &
\colhead{f([OIII]5007)}    &   
\colhead{f([OIII]4959)}    &
\colhead{f([OIII]4363)}    &     
\colhead{f([OII]3727)}  &          
\colhead{$T_e$[OIII]} &
\colhead{12$+$log(O/H)} }

\startdata
\noalign{\vskip-2pt}

 [OIII] emitters & & & & & & \\

 31 &  513.6 $\pm$ 24.0 &  222.3 $\pm$ 11.4 &    $<$ 6.60 &   54.4 $\pm$  4.62 &  $<$ 1.19  &  $>$ 8.09  \\ 
  
 40 &  577.9 $\pm$ 21.6 &  191.3 $\pm$ 8.05 &    9.40 $\pm$  3.40 &  140.9 $\pm$  6.22 &  1.14 $<$  1.34 $<$  1.53 & 7.86 $<$ 8.03 $<$ 8.25 \\ 
  
 51 &  401.5 $\pm$ 12.6 &  146.9 $\pm$  5.52 &    9.40 $\pm$  4.50 &   $<$  2.39  &  1.19 $<$  1.55 $<$  1.90 & 7.51 $<$ 7.62 $<$ 7.94 \\

 76 &  464.4 $\pm$ 10.5 &  191.3 $\pm$  4.86 &    $<$ 2.90  &  344.5 $\pm$  8.07 &  $<$  0.95  & $>$ 8.55  \\ 
 
118 &  492.6 $\pm$ 29.7 &  193.9 $\pm$ 12.9 &   34.4 $\pm$ 12.8 &   11.3 $\pm$  2.61 &  2.16 $<$  3.08 $<$  4.32 & 6.93 $<$ 7.16 $<$ 7.44 \\

195 &  335.0 $\pm$ 21.4 &  129.5 $\pm$ 10.2 &   24.0 $\pm$ 10.9 &   97.1 $\pm$  9.71 &  2.02 $<$  3.17 $<$  4.86 & 6.78 $<$ 7.06 $<$ 7.44 \\

206 &  597.1 $\pm$ 19.5 &  204.1 $\pm$  7.41 &   21.6 $\pm$  9.20 &   $<$ 1.72  &  1.48 $<$  1.97 $<$  2.48 & 7.42 $<$ 7.55 $<$ 7.84 \\

208 &  658.0 $\pm$ 30.9 &  249.8 $\pm$ 12.3 &   15.6 $\pm$  8.00 &   $<$ 22.1  &  1.16 $<$  1.56 $<$  1.93 & 7.67 $<$ 7.85 $<$ 8.22 \\

223 &  242.9 $\pm$ 15.3 &   83.3 $\pm$  7.53 &   23.7 $\pm$ 18.9 &   $<$ 10.1  &  1.45 $<$  4.64 $<$ 19.62 & 6.14 $<$ 6.61 $<$ 7.53 \\

252 &  466.8 $\pm$  9.32 &  157.9 $\pm$  3.57 &    9.20 $\pm$  4.00 &  139.0 $\pm$  3.71 &  1.16 $<$  1.45 $<$  1.72 & 7.68 $<$ 7.87 $<$ 8.14 \\

\enddata
\tablecomments{Only emitters with $>15\sigma$ H$\beta$ fluxes are listed. All fluxes are normalized by their f(H$\beta$) 
and multiplied by 100. 1$\sigma$ upper limits are listed for [OII]3727 flux below 3$\sigma$ and [OIII]4363 below 
1$\sigma$.  The units of $T_e$[OIII] are $10^{-4}$ [K].}
\end{deluxetable}

  \begin{figure*}
  \begin{centering}
  \figurenum{17}
   \includegraphics[height=6.0in,clip,angle=270,scale=1.0]{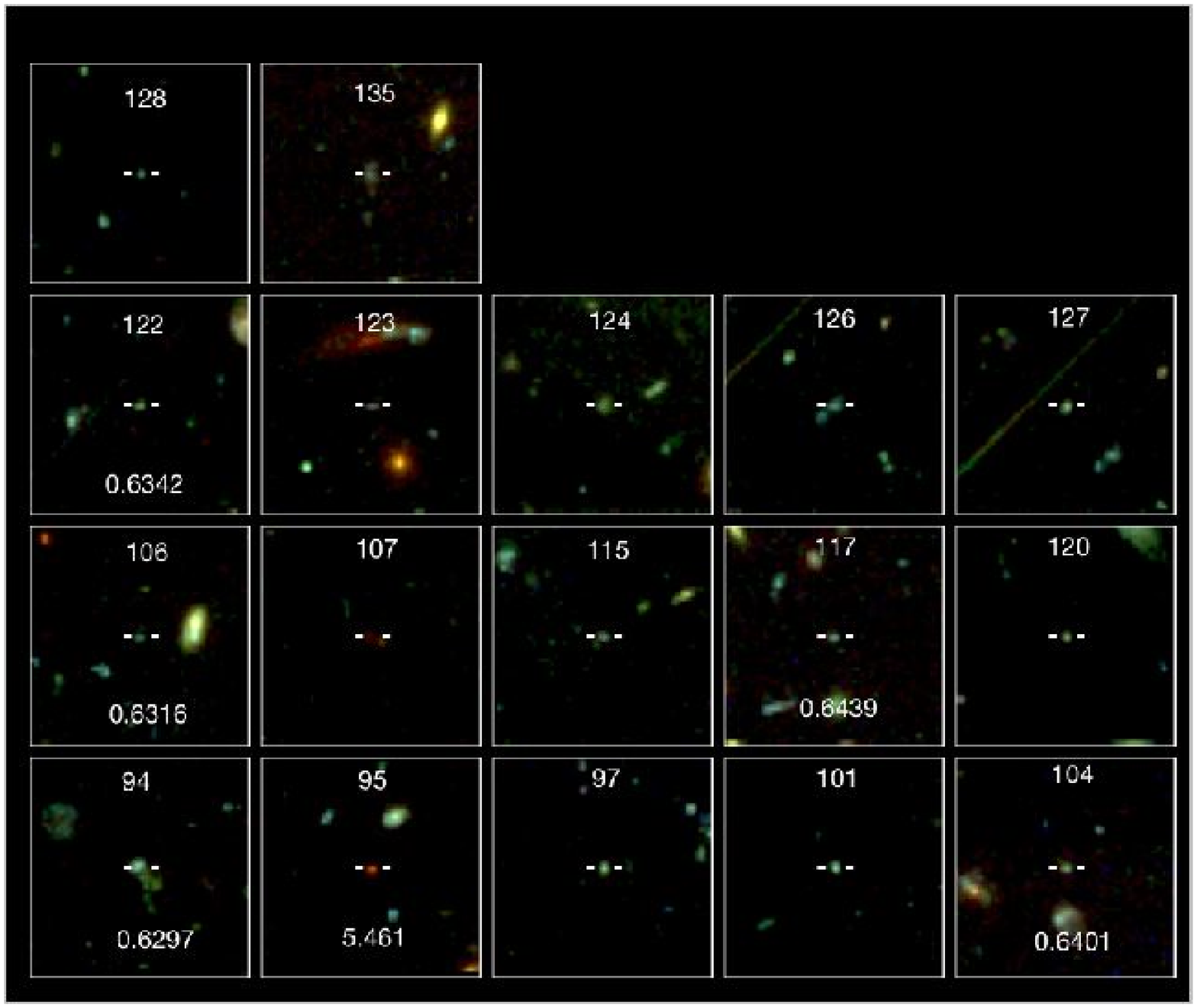}  
  \caption{HST/ACS (B, V, z') composite images of NB816 emitters in the GOODS-N field with overlaid object IDs from Table~\ref{tbl-2} and redshifts, where known. Fields are $12\farcs5$ on a side.
  \label{figy:hst_816}}
  \end{centering}
  \end{figure*}

%
\begin{deluxetable}{ccccccc}
\tablecaption{Line Fluxes and Oxygen Abundances for L912 selected emitters\label{tbl-5}}
\tablewidth{0pt}
\scriptsize

\tablehead{
\colhead{Object}           &
\colhead{f([OIII]5007)}    &   
\colhead{f([OIII]4959)}    &
\colhead{f([OIII]4363)}    &     
\colhead{f([OII]3727)}  &          
\colhead{$T_e$[OIII]} &
\colhead{12$+$log(O/H)} }

\startdata
\noalign{\vskip-2pt}

 [OIII] emitters & & & & & & \\
 
  3 &  550.9 $\pm$ 12.9 &  187.9 $\pm$  4.91 &   23.9 $\pm$  7.90 &    8.6 $\pm$  2.5 & 1.74 $<$ 2.20 $<$ 2.71 & 7.26 $<$ 7.43 $<$ 7.65 \\

  6 &  588.1 $\pm$ 35.1 &  216.0 $\pm$ 14.2 &   18.4 $\pm$ 11.0 &   52.0 $\pm$  8.6 & 1.20 $<$ 1.79 $<$ 2.39 & 7.40 $<$ 7.68 $<$ 8.14 \\

  9 &  442.3 $\pm$ 15.3 &  154.7 $\pm$ 6.42 &   $<$ 12.3        &  157.2 $\pm$  6.88 & $<$ 1.70  & $>$ 7.70   \\

 10 &  490.0 $\pm$ 11.9 &  178.7 $\pm$  5.14 &   13.7 $\pm$  4.40 &    $<$ 2.95        & 1.42 $<$ 1.69 $<$ 1.97 & 7.55 $<$ 7.61 $<$ 7.82 \\

 17 &  342.5 $\pm$ 20.0 &  129.7 $\pm$  9.29 &   15.9 $\pm$  9.40 &    $<$ 7.72        & 1.41 $<$ 2.26 $<$ 3.28 & 7.03 $<$ 7.22 $<$ 7.70 \\

 20 &  418.7 $\pm$ 17.4 &  135.1 $\pm$  6.96 &   16.8 $\pm$  5.50 &   24.5 $\pm$  2.45 & 1.69 $<$ 2.11 $<$ 2.57 & 7.18 $<$ 7.36 $<$ 7.58 \\

239 &  202.4 $\pm$ 10.2 &   75.6 $\pm$  6.40 &   $<$ 8.20         &  190.6 $\pm$ 10.2 &  $<$ 2.08  & $>$ 7.34  \\

270 &  351.7 $\pm$ 15.1 &  149.7 $\pm$  7.73 &   12.4 $\pm$  3.40 &   30.7 $\pm$  2.72 & 1.59 $<$ 1.87 $<$ 2.16 & 7.28 $<$ 7.43 $<$ 7.61 \\

\hline

 H$\alpha$ emitters  & &  & & &  & \\
 52 &  589.1 $\pm$ 10.0 &  179.2 $\pm$  3.42 &   18.3 $\pm$  1.59 &   $<$ 1.56         & 1.75 $<$ 1.83 $<$ 1.92 & 7.62 $<$ 7.67 $<$ 7.72 \\ 
 60 &  619.1 $\pm$ 33.5 &  206.7 $\pm$ 12.5 &   10.7 $\pm$  7.77 &   48.4 $\pm$  8.7 & 0.90 $<$ 1.37 $<$ 1.77 & 7.67 $<$ 7.96 $<$ 8.57 \\ 
266 &  682.8 $\pm$ 10.3 &  217.7 $\pm$  3.57 &   14.7 $\pm$  2.42 &    ...            & 1.40 $<$ 1.52 $<$ 1.63 & $<$ 8.3   \\  

\enddata

\tablecomments{Same as Table~\ref{tbl-4} but for NB912 emitters. The [OII]$\lambda$3727 of ID266 is beyond
the DEIMOS wavelength coverage and thus was not being measured.  }
\end{deluxetable}

  \begin{figure*}
  \begin{centering}
  \figurenum{18}
  \includegraphics[height=6.0in, clip,angle=270,scale=1.0]{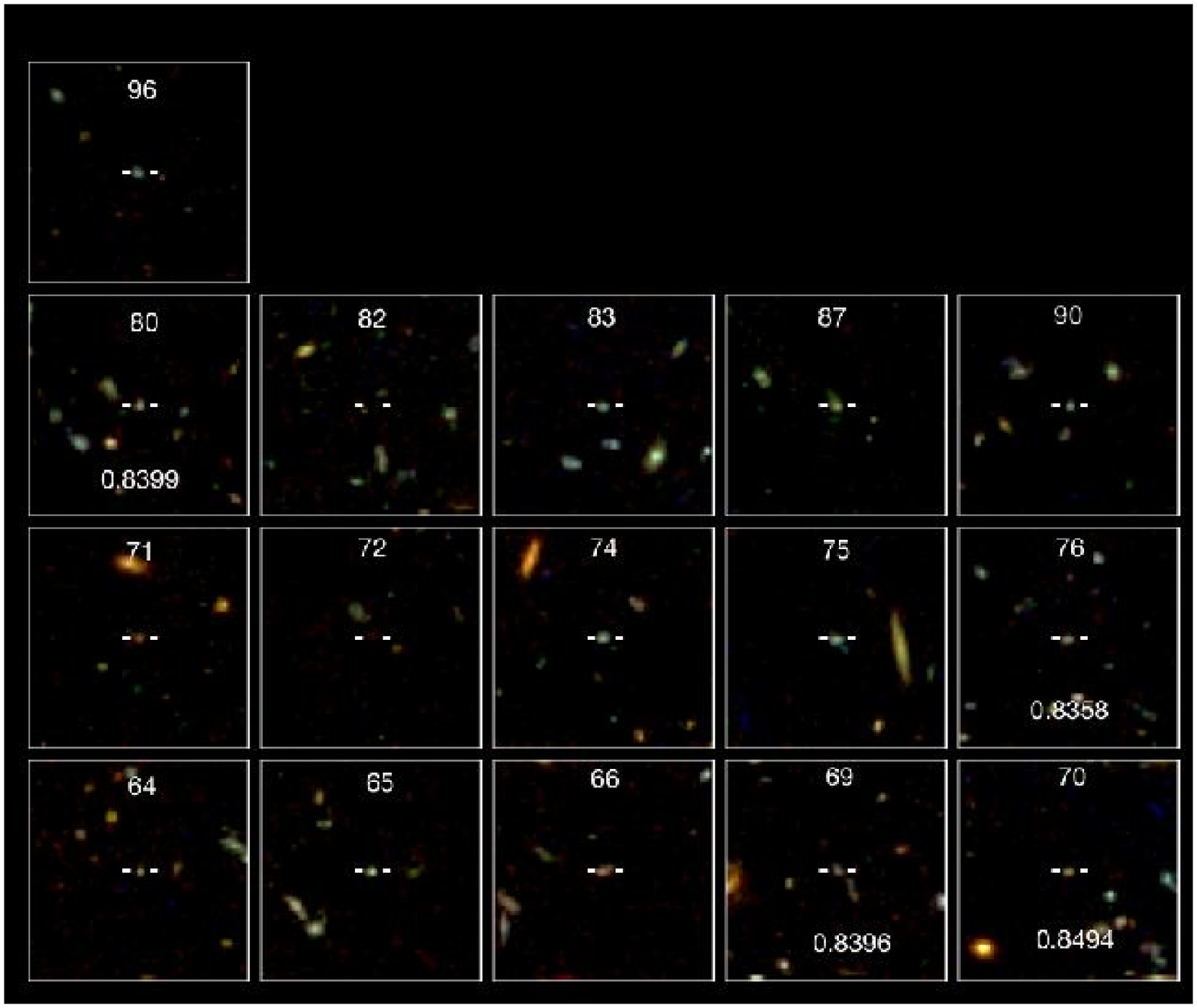}
  \caption{HST/ACS (B, V, I) composite images of NB912 emitters in the GOODS-N field with overlaid object IDs from Table~\ref{tbl-3} and redshifts, where known.  Fields are $12\farcs5$ on a side.
  \label{figz:hst_912}}
  \end{centering}
  \end{figure*}

\end{document}